\documentclass[aps, prl,
reprint,
superscriptaddress,
floatfix
]{revtex4-1}
\usepackage{graphicx}
\begin{document}
	\title{Integer quantum Hall transition on a tight-binding lattice}
	\def\afTUC{Institute of Physics, Chemnitz University of Technology, 09107 Chemnitz, Germany}
	\def\afMST{Department of Physics, Missouri University of Science and Technology, Rolla, Missouri 65409, USA}
	\author{Martin Puschmann}
	\email[]{puschmannm@mst.edu}
%	\affiliation{\afTUC}
	\affiliation{\afMST}
	\author{Philipp Cain}
	\affiliation{\afTUC}
	\author{Michael Schreiber}
	\affiliation{\afTUC}
	\author{Thomas Vojta}
	\affiliation{\afMST}
	
	\date{\today}
	
	\begin{abstract}
		Even though the integer quantum Hall transition has been investigated for nearly four decades its critical behavior remains a puzzle. The best theoretical and experimental results for the localization length exponent $\nu$ differ significantly from each other, casting doubt on our fundamental understanding. While this discrepancy is often attributed to long-range Coulomb interactions, Gruzberg et al. [Phys.\,Rev.\,B \textbf{95}, 125414 (2017)] recently suggested that the semiclassical Chalker-Coddington model, widely employed in numerical simulations, is incomplete, questioning the established central theoretical results. To shed light on the controversy, we perform a high-accuracy study of the integer quantum Hall transition for a microscopic model of disordered electrons. We find a localization length exponent $\nu=2.58(3)$ validating the result of the Chalker-Coddington network.
	\end{abstract}
	
	\maketitle
\begin{figure}
		\centering
		\includegraphics{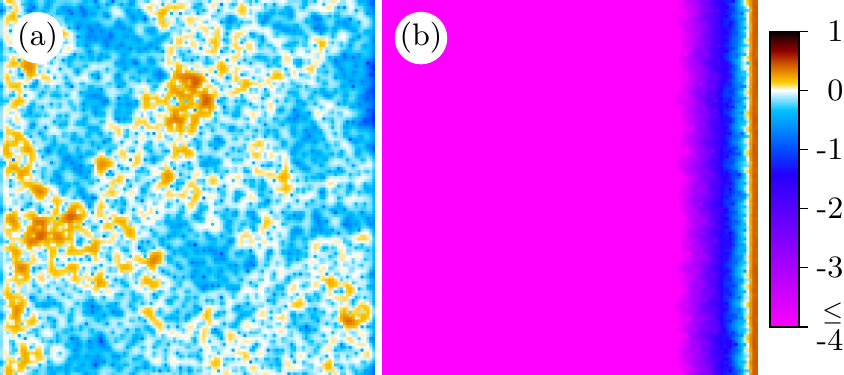}
		\caption{Critical state (left, $E=E_\mathrm{c}=3.4221$) and edge state (right, $E=3.0056$) for a square system of linear size $L=128$, magnetic flux $\Phi=1/10$ and disorder strength $W=0.5$. Coloring represents wave function intensities $|\psi_j|^2$ at site $j$ via $\log_{L^2}(L^2|\psi_j|^2$). Periodic and open boundary conditions are applied in $y$ and $x$ direction, respectively.}
		\label{fig:eigenstates}
	\end{figure}

	%%%%%%%%%%%%%%%%%%%%%%%%%%%%%%%%%%%%%%%%%%%%%%%%%%%%%%%%%%%%%%%%%%%%%%%%%%%%%%%%%%%
	% Introduction
	%%%%%%%%%%%%%%%%%%%%%%%%%%%%%%%%%%%%%%%%%%%%%%%%%%%%%%%%%%%%%%%%%%%%%%%%%%%%%%%%%%%
	The discovery of the integer quantum Hall (IQH) effect by Klitzing \textit{et al.} \cite{KliDP80} opened a new area in condensed matter physics, combining topology and Anderson localization. Two-dimensional electrons in a perpendicular magnetic field $B$ follow circular cyclotron orbits that are quantized into discrete \textit{Landau levels} (LLs), having energies $E_n=(n+1/2)\hbar\omega$, with integer $n$ and cyclotron frequency $\omega=\mathrm{e} B /\mathrm{m}$. Disorder broadens each LL into a Landau band (LB) and localizes all electronic states in the bulk except for one critical state in each band center (Fig.\ \ref{fig:eigenstates}a).  At the system rim, skipping cyclotron orbits lead to edge states (Fig.\ \ref{fig:eigenstates}b) that are extended along the boundary even in the presence of disorder and cause the plateau structure of the Hall conductance. Quantum Hall states are thus examples of topological insulators, bulk insulators with protected edge states. When the Fermi energy is tuned through one of the critical states, the system undergoes a quantum phase transition between two localized phases. This IQH transition belongs to the realm of Anderson transitions, and the critical exponent $\nu$ describes the divergence $\xi\sim\left|E-E_\mathrm{c}\right|^{-\nu}$ of the localization length $\xi$ at the IQH critical point $E_\mathrm{c}$ \cite{And58,EveM08}.
	
	Critical scaling at the IQH transition has been observed in a number of systems \cite{LiCT05, LiVX09, GieZP09}. Recent experiments on $\mathrm{Al}_x\mathrm{Ga}_{1-x}\mathrm{As}$ heterostructures \cite{LiVX09} measure $\nu$ and the dynamical exponent $z$ independently of each other, giving $\nu=2.38(5)$ and $z=1$.
	While early numerical investigations of the IQH transition provided $\nu$ values in the range of $2.3$ to $2.4$ \cite{HucK90,Huc92,Huc95,LeeW96,HuoB92,CaiRR03,MkhR2009}, several newer high-accuracy studies based on the semiclassical Chalker-Coddington (CC) network model yielded significantly larger values $\nu\approx2.60$ with small error bars of about $0.02$ \cite{ChaC88,SleO09,LiVX2009,SleO12,ObuSF10,AmaMS11,FulHA11,ObuGE12,NudKS15}. The disagreement between the best experimental and theoretical values casts doubt on our fundamental understanding of IQH transitions. How can this exponent puzzle be solved? One candidate is the electron-electron interaction that is not taken into account in the simulations but clearly present in real materials \cite{Huc95,LiVX09,GieZP09,PolS93,PruB95,ZiqFG00,PruB08,BurBE11,Note1}.
%	\footnote{Whereas short-range interactions are irrelevant at the noninteracting fixed point, long-range Coulomb interactions are believed to be relevant \cite{LeeW96}.}
	However, Gruzberg \textit{et al.} \cite{GruKN17} recently questioned the validity of the semiclassical CC network even within the single-particle framework. They suggested that the CC network is too regular and does not contain all types of disorder that are relevant at the IQH transition. For a modified network model, they observed $\nu\approx2.37$ remarkably close to the experimental observations. A recent numerical study of the quantum Hall problem in the presence of $\delta$-impurity potentials gave $\nu=2.4(1)$ \cite{IppSB18} and a Chern number calculation obtained $\nu=2.48(2)$ \cite{ZhuWBW18}, adding to the controversy.
	% A recent numerical study of the quantum Hall problem in the presence of $\delta$-impurity potentials gave $\nu=2.4(1)$ \cite{IppSB18}, adding to the controversy.
	
	In this Letter, we address this controversy by studying the IQH transition in a microscopic tight-binding model of non-interacting electrons on a simple square lattice. We perform a careful finite-size scaling analysis for large systems of up to $768\times 10^6$ lattice sites. In the universal regime, where neither LL coupling nor the nonzero intrinsic LL width affect the transition, we find $\nu=2.58(3)$ in agreement with the (standard) CC network. Thus, the discrepancy between theory and experiment persists even for a microscopic model of non-interacting electrons, pointing to the Coulomb interactions as main culprit.
	
	In the rest of the Letter, we introduce the tight-binding model as well as our numerical method. We then present our results and compare them with those of Refs. \cite{GruKN17} and \cite{IppSB18}.We conclude by discussing broader implications for the IQH transition.
	
	%%%%%%%%%%%%%%%%%%%%%%%%%%%%%%%%%%%%%%%%%%%%%%%%%%%%%%%%%%%%%%%%%%%%%%%%%%%%%%%%%%%%
	% Model and Methods
	%%%%%%%%%%%%%%%%%%%%%%%%%%%%%%%%%%%%%%%%%%%%%%%%%%%%%%%%%%%%%%%%%%%%%%%%%%%%%%%%%%%%
	
	We consider a tight-binding model of non-interacting electrons moving on a square lattice in the presence of a perpendicular magnetic field $\mathbf{B}$. The Hamiltonian, a generalization of the Anderson model of localization, reads
	\begin{equation}
	\mathbf{H}=\sum\limits_{j}u_j\left|j\right\rangle\!\left\langle j\right| + \sum\limits_{\langle j,k \rangle}\exp(i\varphi_{jk})\left|j\right\rangle\!\left\langle k\right|\, .\label{eq:Hamiltonian}
	\end{equation}
	Here, $\left|j\right\rangle$ denotes a Wannier orbital on site $j$. The potentials $u_j$ are independent random variables drawn from a uniform distribution in the interval $[-W/2, W/2]$, characterized by disorder strength $W$. The hopping terms between nearest neighbors $\langle j,k \rangle$ have constant magnitude but complex phase shifts induced by the magnetic field \cite{Pei33,Lutt51}. Choosing the Landau gauge for the vector potential, $\mathbf{A}=(0,Bx,0)$, the Peierls phases read
	\begin{equation}
	\varphi_{jk}=\frac{\mathrm{e}}{\hbar}\int\limits_{j}^{k}\mathbf{A}\cdot\mathrm{d}\mathbf{r}=\cases{
		0 & in $x$ direction \cr
		\pm 2\pi\Phi x_j & in $\pm y$ direction\cr
	}\, ,
	\label{eq:peierlsphase}
	\end{equation}
	where $x_j$ is the $x$ coordinate of site $j$ in units of the lattice constant $l$. $\Phi=Bl^2\mathrm{e}/\mathrm{h}$ is the magnetic flux through a unit cell in units of the flux quantum $\mathrm{h}/\mathrm{e}$.
	
\begin{figure}
		\centering
		\includegraphics{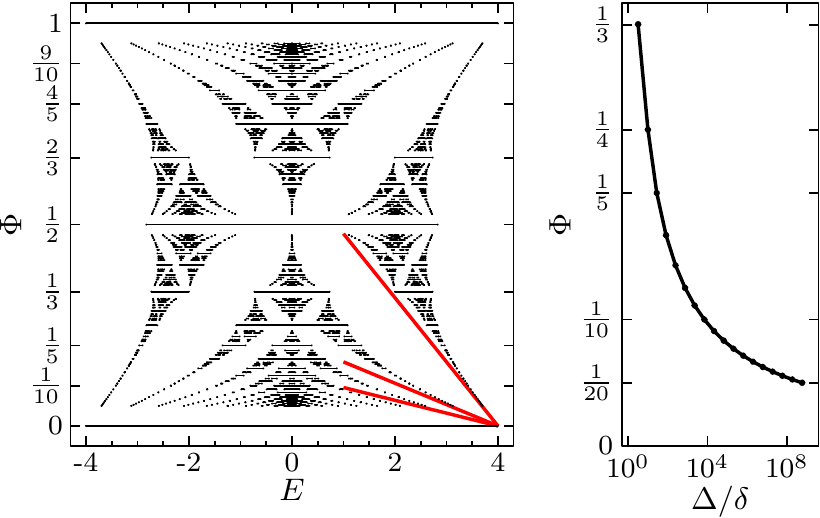}
		\caption{Left: Hofstadter butterfly showing the energy spectrum of Hamiltonian\ (\ref{eq:Hamiltonian}) in the clean case ($W=0$). For $\Phi=0$ and $1$, the graph shows a single band from $E=-4$ to $4$. Additionally, all Landau levels are shown for $\Phi=p/q$ with coprime numbers $p,q\leq 20$. The free-electron-gas approximation (solid red lines), $E^{+}_n=4 - 4\pi\Phi(n+1/2)$, is shown for $n=0$, $1$, and $2$. Right: LL spacing (mid-band distance $\Delta$ between the LLs with $n=0$ and $n=1$) in multiples of the intrinsic LL width $\delta$ as function of $\Phi$.}
		\label{fig:butterfly}
	\end{figure}

	In the absence of disorder, $W=0$, the energy spectrum of the Hamiltonian (\ref{eq:Hamiltonian}) is fractal and takes the form of the famous Hofstadter butterfly \cite{Hof76,Ram85}, visualized in Fig.\ \ref{fig:butterfly}. 
	The spectrum is symmetric with respect to energy $E=0$ reflecting the bipartite character of the lattice. The spectrum is also periodic in $\Phi$ with period $1$ and symmetric with respect to $\Phi=1/2$. For a rational $\Phi=p/q$ with coprime integers $p$ and $q$, the spectrum splits into exactly $q$ LLs \cite{Note2}.
%	\footnote{For irrational $\Phi$, in contrast, the spectrum consists of dense but isolated eigenvalues.}.
	In contrast to the free-electron gas, each LL is a continuous band, having a \emph{nonzero intrinsic width}. 
	In the limit of small $\Phi$ and for small LL indices $n$, the spectra become similar to LLs of a free-electron gas; i.e., their widths go to zero and their energies follow $E^{\pm}_n=\pm 4 \mp 4\pi\Phi(n+1/2)$ where $\pm$ distinguishes the positive and negative sides of the spectra. 
	
	Random potentials, $W>0$, broaden the LLs further, and all extended states turn into localized states except for one critical state in each LL that shows multifractal fluctuations. To determine the universal properties of the IQH transition, we need to consider parameters for which neither LL coupling nor the intrinsic LL width play a role. The disorder must thus lead to LL broadening that is much larger than the intrinsic LL width $\delta$ but much smaller than the LL spacing $\Delta$. The dependence of $\delta$ and $\Delta$ on $\Phi$ is presented in the right panel of Fig.\ \ref{fig:butterfly}, demonstrating that both conditions are easy to fulfill for small $\Phi$, where $\Delta/\delta\gg1$. However, the magnetic length $L_\mathrm{B}=\sqrt{(2n+1)/(2\pi\Phi)}$ becomes very large for small $\Phi$, making the effective size $L/L_\mathrm{B}$ of the system too small. Hence, we expect the numerically most favorable situation for the lowest LL, $n=0$, and moderate $\Phi$.        
	
	We investigate the behavior of the electronic states by means of the recursive Green's function algorithm \cite{SchKM85,SchKM84,KraSM84}. It considers a quasi-one-dimensional strip of $L\times N$ sites ($N\gg L$) which is divided into layers of $L\times 1$ sites. This allows us to calculate the Green's function $\mathbf{G} =((E+i\eta)\mathbf{I}-\mathbf{H})^{-1}$ iteratively layer by layer. The smallest positive Lyapunov exponent $\gamma$ is calculated from the Green's function between the two ends of the strip, i.e., between the layers $x_{j}=1$ and $x_{j}=N$ \cite{Note3},
%	\footnote{For numerical stability, we also calculate $\gamma$ iteratively (in $100$-layer steps) by the algorithm of Refs. \cite{SchKM84,KraSM84}.},
	\begin{eqnarray}
	\gamma=-\lim\limits_{N\rightarrow\infty} \lim\limits_{\eta\rightarrow 0}\frac{\ln\mathrm{tr}\,\left|\mathbf{G}_{1,N}\right|^2}{2N}\quad. 
	\end{eqnarray} 
	We consider strips along the $x$ direction so that periodic boundary conditions can be applied in the layer direction without restricting the value of $\Phi$. To determine the critical behavior, we perform finite-size scaling of the dimensionless Lyapunov exponent $\Gamma\equiv\gamma L$ by varying the strip width $L$. Our results for $\Gamma(E,\Phi,L)$ are ensemble averages of $50$ strip realizations of width $L\leq 512$ and length $N=10^6$ for each $E$ and $\Phi$. For $\Phi=1/10$, we increase the accuracy by using $L\leq 768$ and $150-200$ strips per data point. The relative statistical uncertainties of $\Gamma$ scale approximately with $\sqrt{L}$ and range from $0.0003$ for $L=16$ to $0.0022$ for $L=768$. We use a complex energy shift $\eta=10^{-14}$ to approximate the limit $\eta\rightarrow0$.
	
	%%%%%%%%%%%%%%%%%%%%%%%%%%%%%%%%%%%%%%%%%%%%%%%%%%%%%%%%%%%%%%%%%%%%%%%%%%%%%%%%%%%
	% Results -- Scaling
	%%%%%%%%%%%%%%%%%%%%%%%%%%%%%%%%%%%%%%%%%%%%%%%%%%%%%%%%%%%%%%%%%%%%%%%%%%%%%%%%%%%
	
	Let us now turn to the results. We investigate the transition in the lowest LB for $\Phi=1/3$, $1/4$, $1/5$, $1/10$, $1/20$, $1/100$, and $1/1000$ \cite{Note4}.
%	\footnote{Since the spectrum is symmetric with respect to $E=0$, we consider only the positive side of the spectrum, where the lowest LL appears on the high energy band edge.}.
	(We consider only ratios of the form $1/q$ to avoid splitting of the LB into subbands.) For each $\Phi$, we choose an appropriate disorder strength $W$ such that $\delta<W\lesssim\Delta$. Details of the simulation parameters are provided in the Supplemental Material \cite{Note5}
%	\footnote{See Supplemental Material at XXX for details of the simulation parameters, the finite-size corrections at the critical point, details of the scaling analysis, and transitions in cases of non-negligible intrinsic LL width.}
	, which also shows the density of states in the presence of disorder for several $\Phi$. 
	For reliable results, we analyze the critical behavior of the Lyapunov exponent $\Gamma$ in two different ways: First, we analyze the energy and system-size dependencies of $\Gamma$ graphically. Then we apply compact, sophisticated scaling functions to our data. 
	
\begin{figure}
		\includegraphics{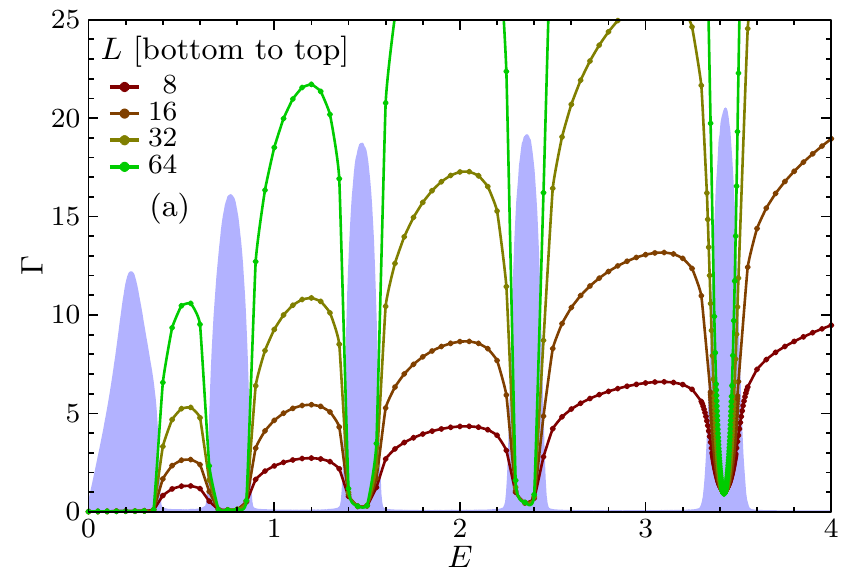}
		\includegraphics{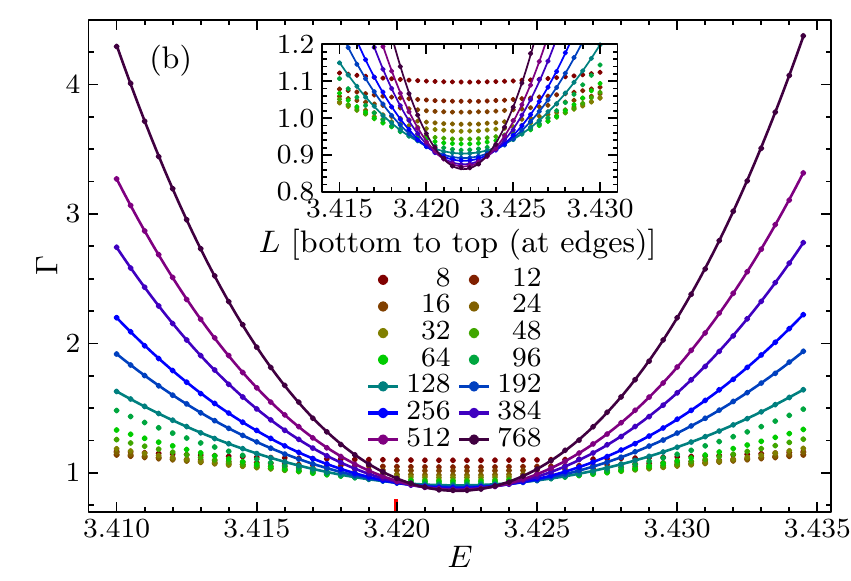}
		\caption{(a)~Dimensionless Lyapunov exponent $\Gamma(E,L)$ for $\Phi=1/10$ and $W=0.5$ as function of $E$ for several $L$. The statistical errors are well below the symbol size. The blue area shows the density of states in arbitrary units. Lines are guides to the eye only. (b)~$\Gamma(E,L)$ close to the lowest IQH transition. The solid lines are fits (see main text for details). The inset shows the immediate vicinity of the transition. The tiny red bar at the bottom shows the energy range of the LL in absence of disorder (compare Fig S1).}
		\label{fig:transition}
	\end{figure}

	The location of the transition, i.e., the critical energy $E_\mathrm{c}$, is identified as the position of the minimum of $\Gamma$ as function of $E$ (see Fig.\ \ref{fig:transition}).
	\begin{figure}
		\centering
		\includegraphics{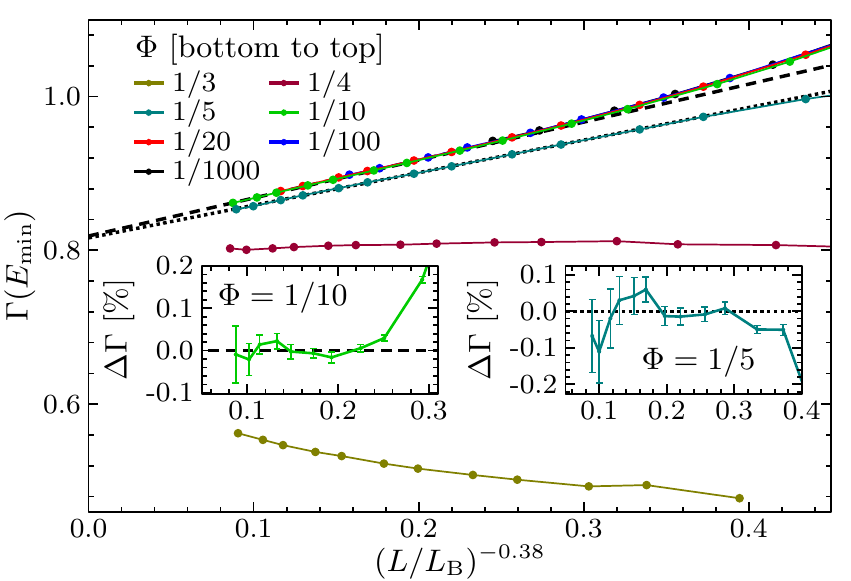}
		\caption{System-size dependence of $\Gamma$ at $E_{\text{min}}$ for several $\Phi$ under assumption of an irrelevant exponent $y=0.38$. The fits of eq.\ (\ref{eq:critical_scaling}) for $\Phi\leq1/10$ (black dashed line) and $\Phi=1/5$ (black dotted line) are based on the data with $L\geq128$. The statistical errors are below the symbol size. The insets show the deviations $\Delta\Gamma$ of the data points from the fit functions. Solid lines are guides to the eye only.}
		\label{fig:critical_behavior}
	\end{figure}     
	Figure \ref{fig:critical_behavior} shows the system-size dependence of $\Gamma$ at the minimum position $E_\mathrm{min}$ for several $\Phi$ (where $E_\mathrm{min}\rightarrow E_\mathrm{c}$ for $L\rightarrow\infty$). Expressing the strip width $L$ in multiples of the magnetic length $L_\mathrm{B}=1/\sqrt{2\pi\Phi}$, all data line up for small flux $\Phi\lesssim 1/10$. We conclude that these data are unaffected by LL coupling and the nonzero LL width and therefore represent the universal behavior of the IQH transition. For $\Phi=1/5$, the significantly increased intrinsic LL width, $\Delta/\delta\approx 28$, leads to deviations, but the asymptotic behavior seems to be similar to the universal case. For higher $\Phi$, the modifications are much stronger, indicating nonuniversal behavior. In the following we therefore focus on the universal regime, $\Phi\lesssim 1/10$.
	
	The strong system-size dependence of the critical Lyapunov exponent implies that corrections to scaling are important. We provide a comprehensive discussion supported by visualization of finite-size corrections in the Supplemental Material \cite{Note5}. In the simplest case, they are described by a power law
	\begin{eqnarray}
	\Gamma(E_\mathrm{c}, \Phi, L)=\Gamma_\mathrm{c}\left(1+a (L/L_\mathrm{B})^{-y}\right) \label{eq:critical_scaling}
	\end{eqnarray}
	with an irrelevant exponent $y>0$. In the literature, the estimates for $y$ vary widely between about $y\approx 0.15$ \cite{SleO09,NudKS15} and $y\approx 0.6\dots0.8$ \cite{ObuGE12,WanLS98,GruKN17}. $y$ values at both ends of this range do not describe our data in the universal regime. However, a mid-range value, $y=0.38$ \cite{SleO12,ObuGE12,Huc94}, correctly describes the behavior from $L\approx 44 L_\mathrm{B}$ to our largest sizes of $L\approx 608 L_\mathrm{B}$, yielding $\Gamma_\mathrm{c}=0.818(1)$. We also consider the possibility that the leading corrections to scaling take a logarithmic form $\Gamma=\Gamma_\mathrm{c}[1+a/\log(\lambda (L/L_\mathrm{B}))]$ (with unknown scaling factor scale $\lambda$) \cite{AmaMS11,NudKS15}. Our universal data ($\Phi\leq 1/10$) for $L\geq 20L_\mathrm{B}$ are well described by the logarithmic form, giving $\Gamma_\mathrm{c}=0.738(1)$ and $\lambda=1.68(4)$.
	
	%%%%%%%%%%%%%%%%%%%%%%%%%%%%%%%%%%%%%%%%%%%%%%%%%%%%%%%%%%%%%%%%%%%%%%%%%%%%%%%%%%%
	% Results -- Transition
	%%%%%%%%%%%%%%%%%%%%%%%%%%%%%%%%%%%%%%%%%%%%%%%%%%%%%%%%%%%%%%%%%%%%%%%%%%%%%%%%%%%
	
	Besides $\Gamma(E_\mathrm{c})$, we also look at the curvature $\Gamma''(E_\mathrm{c})$ which is supposed to scale as $L^{2/\nu}$ in the large $L$ limit. Figure~S3 in \cite{Note5} shows the curvature as function of $L/L_\mathrm{B}$. Again, all data for $\Phi\lesssim 1/10$ line up, confirming the  universal regime. Moreover, the graphical analysis provides strong evidence for the asymptotic value of $\nu$ to be significantly larger than $2.4$.
	
	For reliable quantitative estimates, we now focus on fits of a sophisticated scaling function $\Gamma(x_{\text{r}} L^{1/\nu}, x_{\text{i}} L^{-y})$ that provide combined estimates for all critical parameters, i.e., $\Gamma_\mathrm{c}$, $\nu$, and $y$, simultaneously. In the Supplemental Material \cite{Note5}, we describe in detail this function and its expansion in terms of the relevant scaling field $x_{\text{r}} L^{1/\nu}$ and irrelevant scaling field $x_{\text{i}} L^{-y}$ with scaling variables $x_{\text{r}}(E-E_\mathrm{c})$ and $x_{\text{i}}(E-E_\mathrm{c})$.   
	
	Based on the above discussion, a flux of $\Phi=1/10$ is most suitable to obtain high-accuracy critical parameters for the IQH transition, because (i) the data fall onto the universal master curve in Fig.\ \ref{fig:critical_behavior} and (ii) we can still reach large effective sizes up to $L=608 L_\mathrm{B}$. 
	
	Figure\ \ref{fig:transition}b shows $\Gamma$ with $50$ energy points in vicinity of $E_\mathrm{c}$ for each $L$. We obtain the best fits if we neglect smaller systems, $L<128$, and include the leading order in terms of the irrelevant scaling field only. This yields an IQH transition at $E_\mathrm{c}=3.422151(3)$ and $\Gamma_\mathrm{c}=0.814(6)$ with the critical exponents $\nu=2.594(14)$ and $y=0.357(26)$. Neglecting further systems leads to consistent results but increased uncertainties. Without irrelevant finite-size corrections, we observe $\nu=2.595(12)$ based on our two largest systems. To extend the fit to smaller systems $64\lesssim L< 128$, one needs to consider higher correction orders, but the reliability of the estimates decays. Based on the robustness of the results with respect to the system-size range and fit expansions, we estimate the critical parameters to be $\Gamma_\mathrm{c}=0.815(8)$, $\nu=2.58(3)$, and $y=0.35(4)$, as explained in \cite{Note5}.     
	
	We also perform sophisticated fits for the logarithmic correction-to-scaling scenarios. Employing an irrelevant scaling field $x_\mathrm{i}/(b+x_\mathrm{i}\log L)$ in our scaling ansatz for $\Gamma$ leads to $\Gamma_\mathrm{c}=0.745(6)$, $\nu=2.597(27)$ for $L\geq128$. The inclusion of smaller systems requires higher order corrections, giving $\Gamma_\mathrm{c}\approx 0.74$ and $\nu\approx2.60$ with increased uncertainties. 
	
	For fluxes $\Phi<1/10$, the maximum effective sizes $L/L_\mathrm{B}$ are smaller, leading to more pronounced finite-size effects.
	If higher expansion orders of the irrelevant scaling field are included in the fit, we still find values $\nu\approx2.56$ and $y\approx 0.38$ for $\Phi=1/20$ and $1/100$. However, for $\Phi=1/1000$, our largest system size $L=512$ corresponds to $41 L_\mathrm{B}$ only, and our critical parameters, i.e., $\Gamma_\mathrm{c}\approx0.89$, $\nu\approx2.35$ and $y\approx0.7$, differ significantly from the above values. This clearly shows that a system size of $41 L_\mathrm{B}$ is too small to describe the asymptotic behavior.
	
	Let us briefly comment on the transition for fluxes $\Phi>1/10$, where the intrinsic LL width is comparable with the LL spacing. These transitions are not expected to be in the universal regime. Still, for $\Phi=1/5$, the asymptotic scaling behavior seems to be very similar to that of the universal curve (see Fig.\,\ref{fig:critical_behavior}), and our finite-size scaling analyses provide compatible results. However, for higher $\Phi$, deviations are much stronger. More insight is provided in the Supplemental Material \cite{Note5}. 
	
	%%%%%%%%%%%%%%%%%%%%%%%%%%%%%%%%%%%%%%%%%%%%%%%%%%%%%%%%%%%%%%%%%%%%%%%%%%%%%%%%%%%
	% Conclusion
	%%%%%%%%%%%%%%%%%%%%%%%%%%%%%%%%%%%%%%%%%%%%%%%%%%%%%%%%%%%%%%%%%%%%%%%%%%%%%%%%%%%
	
	In summary, we have investigated the IQH transition in the lowest LL of a microscopic model of non-interacting electrons. In the universal regime where neither the nonzero intrinsic width of the LLs nor LL coupling play a role, we find a localization length exponent of $\nu=2.58(3)$, independent of the details of the finite-size scaling analysis. Based on our numerical data, we are unable to discriminate between power-law corrections to scaling with an irrelevant exponent $y=0.35(4)$ and logarithmic scaling corrections.
	
	Note that our value for the localization length exponent agrees well with recent high-accuracy investigations of the IQH transition in the standard CC model \cite{SleO09,LiVX2009,SleO12,ObuSF10,AmaMS11,ObuGE12,NudKS15}. Our estimates for the critical Lyapunov exponent depend on the correction type: we found $\Gamma_\mathrm{c}=0.815(8)$ and $\Gamma_\mathrm{c}=0.745(6)$ for the power-law and logarithmic corrections, respectively. In both cases, our estimates are slightly larger (by about $0.025$) than those of transfer-matrix calculations on the CC network model \cite{SleO09,AmaMS11,NudKS15}. However, using $\Gamma_\mathrm{c}=\pi(\alpha_0-2)$ \cite{Jan98}, $\Gamma_\mathrm{c}=0.815(5)$ corresponds to the multifractal exponent $\alpha_0=2.2594(15)$ which coincides with $\alpha_0=2.2596(4)$ \cite{EveMM08} and $2.2617(6)$ \cite{ObuSF08} found for the CC network model. 
	
	Our results suggest that the discrepancy between the experimental and theoretical values of $\nu$ cannot be attributed to a too regular structure of the semiclassical CC network model \cite{Note6}.
%	\footnote{The lattice in the CC network plays a very different role than in our work. In the CC network, it represents the semiclassical electron path whereas in our work, it sets up the Hamiltonian while the electron motion is irregular.}.
	Why did the modified CC network \cite{GruKN17}, the Chern number calculation \cite{ZhuWBW18}, and the study of the IQH transition in the presence of $\delta$ impurities \cite{IppSB18} lead to different critical behavior with smaller $\nu$? The latter two investigation used linear system sizes up to about $100 L_\mathrm{B}$, much smaller than our largest sizes of more than $600L_\mathrm{B}$. Moreover, corrections to scaling were not included in the analysis of Ref.\ \cite{IppSB18} or are apparently almost negligible in Ref.\ \cite{ZhuWBW18,Note7}. 
%	\footnote{It is remarkable that the irrelevant exponent $y=4.3(2)$ observed in Ref.\ \cite{ZhuWBW18}, expected to be universal, is much larger than those found in the current work or in other recent investigations based on CC network models.}.
	We therefore believe that the system sizes may be too small to reach the asymptotic regime. (In our system the crossover to the asymptotic behavior occurs for $L\gtrsim 110 L_\mathrm{B}$.) Insufficiently small system sizes are also assumed to be the reason why early numerical studies gave $\nu$ values in the range of $2.3$ to $2.4$ \cite{Note8},
%	\footnote{It is worth emphasizing that simple manual scaling plots are not able to resolve the slowly decaying corrections in this problem.}
	see Fig.~S3 in \cite{Note5}.
	
	The case of the disordered CC network is less clear, as the system sizes in Ref.\ \cite{GruKN17} are comparable to sizes used in recent studies of the standard CC network. Does the deviation between $\nu\approx2.37$ found in Ref.\ \cite{GruKN17} and the value $\nu\approx2.60$ in our model and the standard CC network mean that they belong to different universality classes? It is worth noting that the modified CC network of Ref.\ \cite{GruKN17} has much stronger corrections to scaling (as manifested in a much stronger size dependence of $\Gamma(E_\mathrm{c})$). This may push the crossover to the asymptotic regime to larger sizes. Establishing the universality class of the modified CC model remains a task for the future.
	
	In conclusion, our results imply that the paradigmatic CC network model correctly captures the physics of disordered non-interacting electrons close to the IQH transition, in contrast to what was suggested in Ref.\ \cite{GruKN17}. Electron-electron interactions thus remain the likely culprit for the puzzling disagreement between the best theoretical and experimental results for the critical behavior. Whereas screened interactions were shown to be irrelevant at the noninteracting fixed point, long-ranged Coulomb interactions are believed to be relevant \cite{LeeW96}. However, quantitative results for the critical behavior in the presence of Coulomb interactions do not exist.     
	
	\begin{acknowledgments}
		This work was supported by the NSF under Grant Nos. DMR-1828489, DMR-1506152, PHY-1607611, and PHY-1125915. We thank Ferdinand Evers, Ilya Gruzberg, and Ara Sedrakyan for helpful discussions. T.V. is grateful for the hospitality of the Kavli Institute for Theoretical Physics, Santa Barbara, and the Aspen Center for Physics, where part of the work was performed.
	\end{acknowledgments}
	

%%%%%%%%%%%%%%%%%%%%%%%%%%%%%%%%%%%%%%%%%%%%%%%%%%%%%%%%%%%%%%%%%%%%%%%%%%%%%%%%%%%%%%%%%%%%%%%%%%%%%%%%
\clearpage
%%%%%%%%%%%%%%%%%%%%%%%%%%%%%%%%%%%%%%%%%%%%%%%%%%%%%%%%%%%%%%%%%%%%%%%%%%%%%%%%%%%%%%%%%%%%%%%%%%%%%%%%
% merge with supplemental material

\onecolumngrid
\begin{center}
{\large\bf Supplemental Material for\\
Integer quantum Hall transition on a tight-binding lattice}\\[1em]
{Martin Puschmann,$^{1,*}$ Philipp Cain,$^2$  Michael Schreiber,$^2$  and Thomas Vojta $^1$}\\
$^1$\textit{\small Department of Physics, Missouri University of Science and Technology, Rolla, Missouri 65409, USA}\\
$^2$\textit{\small Institute of Physics, Chemnitz University of Technology, 09107 Chemnitz, Germany}\\
(Dated:\ \today)
\end{center}

\bigskip
\twocolumngrid

\setcounter{equation}{0}
\setcounter{figure}{0}
\setcounter{table}{0}
\setcounter{page}{1}
\makeatletter
\renewcommand{\theequation}{S\arabic{equation}}
\renewcommand{\thefigure}{S\arabic{figure}}
\renewcommand{\bibnumfmt}[1]{[S#1]}
\renewcommand{\citenumfont}[1]{S#1}

In the following sections, we present some additional details about the simulation parameters, the finite-size corrections at the critical point, the details of the finite-size scaling in vicinity of the transition, and the properties of transitions in the case of non-negligible intrinsic Landau level (LL) width.

\section{Details of the simulation parameters}
The Hofstadter butterfly (Fig.\ 2 of the main text) describes the energy spectrum of the tight-binding Hamiltonian (1) in the clean case ($W=0$) as function of the dimensionless magnetic flux $\Phi$. For the investigation of universal properties, we consider the lowest LL, $n=0$, at the positive energy band edge for several $\Phi$.

Table\ \ref{tab:parameter} lists the detailed parameters. 
\begin{table}[b]
	\caption{Parameters of the considered IQH transitions in the lowest LB for several $\Phi$. Based on the Hofstadter butterfly (Fig. 2 of the main text), we identified the intrinsic LL width $\delta$ and the LL separation $\Delta$ between LLs with $n=0$ and $n=1$ (for $\Phi<1/20$, $\Delta\approx4\pi\Phi$). We consider systems of size up to $L_\mathrm{max}$ ($L_\mathrm{B}=1/\sqrt{2\pi \Phi}$ is the magnetic length). The energy value $E_\mathrm{c}$ of the critical point in the lowest LB and its standard deviation result from our finite-size scaling analysis of the data for the dimensionless Lyapunov exponent.}
	\label{tab:parameter}
	\begin{ruledtabular}
		\begin{tabular}{lllllrr}
			\multicolumn{1}{c}{$\Phi$} & \multicolumn{1}{c}{$\delta$} & \multicolumn{1}{c}{$\Delta$} & \multicolumn{1}{c}{$W$} &  \multicolumn{1}{c}{$E_\mathrm{c}$} & \multicolumn{1}{c}{$L_\mathrm{max}$} & \multicolumn{1}{c}{$\frac{L_\mathrm{max}}{L_\mathrm{B}}$} \\ 
			\colrule 
			1/3 & 0.73 & 2.27 & 2.0 & 2.130(2) & 384 & 556 \\
			1/4 & 0.22 & 2.18 & 2.0 & 2.6735(9) & 512 & 642\\
			1/5 & 0.064 & 1.81 & 2.0 & 2.93316(3) & 512 & 573\\
			1/10 & 0.00015 & 1.07 & 0.5 & 3.422151(3) & 768 & 608\\
			1/20 &  \multicolumn{1}{r}{$10^{-9}$} & 0.58 & 0.01 & 3.6980313(3) & 512 & 287 \\
			1/100 & $\ll10^{-9}$ & $0.126$ & 0.01 & 3.9376627(3) & 512 & 128 \\
			1/1000 & $\ll10^{-9}$ & $0.0126$ & 0.001 & 3.993721786(2) & 512 & 41 \\
		\end{tabular}
	\end{ruledtabular}
\end{table}
For $W=0$, the lowest LL has a nonzero intrinsic width $\delta$ and a separation $\Delta$ from the next LL with $n=1$. Both $\delta$ and $\Delta$ depend strongly on $\Phi$. Uniformly distributed random potentials lead to an asymmetric broadening of the LL into LBs. (The total broadening remains less than $W$.) For $\Delta/\delta\gg 1$, we expect universal behavior for a broad range of disorder strength $W$ with $\delta<W\lesssim\Delta$. Note that $E_\mathrm{c}$ is nonuniversal and increases with increasing $W$. In contrast, if $\delta$ becomes significant in comparison to $\Delta$, nonuniversal behavior may emerge. \\

\begin{figure}
	\includegraphics{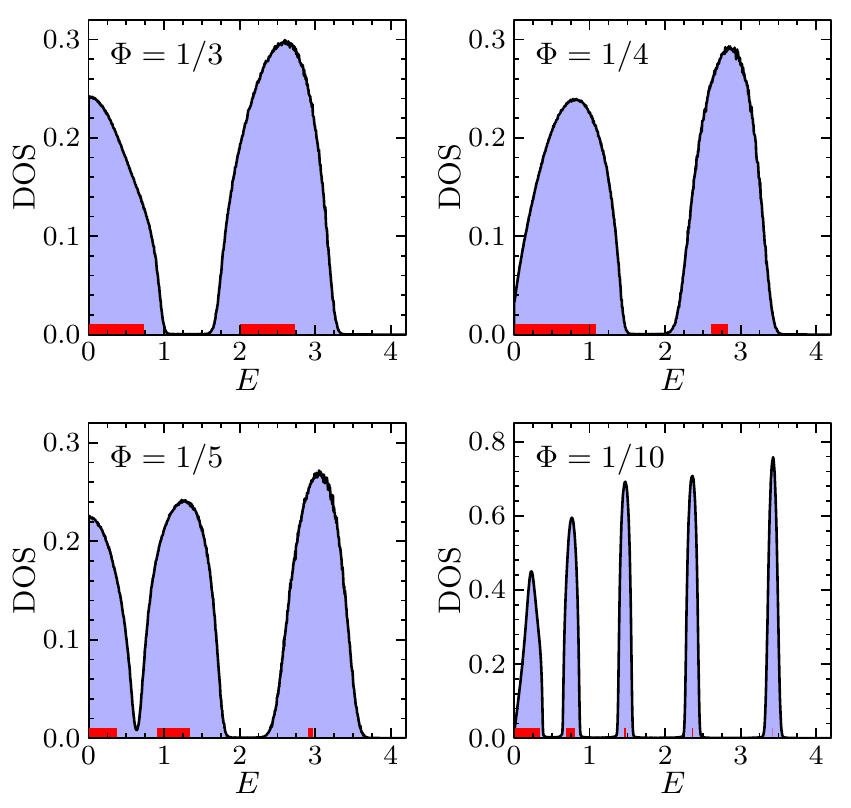}
	\caption{Normalized density of states for $\Phi=1/3$, $1/4$, $1/5$, and $1/10$ with $W=2$, $2$, $2$, and $0.5$, respectively. The data are averages over $10$ strips of size $1000\times256$. The energy resolution is $0.005$ with $\eta=0.001$. The red bars at the bottom show the energy range of the corresponding LL in the absence of disorder (see Fig. 2 of the main text).}
	\label{fig:square_dos_phi}
\end{figure}

Figure\ \ref{fig:square_dos_phi} shows the density of states in the presence of disorder for several fluxes $\Phi\geq 1/10$. The figure demonstrates that (i) the intrinsic LL width is non-negligible and (ii) the LBs are not well separable for $\Phi>1/10$.
In the recursive Green's function method, $\eta$ should be significantly smaller than the energy-level spacing. (A larger $\eta$ leads to raised values of $\Gamma$.) We note that, based on our parameters, the density of states in vicinity of $E_\mathrm{c}$ for $\Phi=1/1000$ is significantly larger than that for $\Phi=1/10$. The correspondingly reduced energy-level spacing requires smaller $\eta$. In all of our cases, $\eta=10^{-14}$ is sufficient to describe the limit $\eta\rightarrow 0$.

In a similar manner, it is worth discussing the limit $N\rightarrow \infty$ in Eq.\ 3 of the main text. For an efficient parallel calculation on a computer with many cores, we average the result over several strips of large but finite length $N$. To confirm that the statistical error dominates the systematic error due to the finite strip length $N$, we compared the Lyapunov exponents based on $200$ strips of length $N=10^6$ with with those for $20$ strips of length $N=10^7$ for several energy values and a strip width $L=256$ and $\Phi=1/10$. The results agree within their (statistical) errors, confirming a negligible impact on the finite length $N=10^6$ to our results.

\section{Finite-size corrections at the critical point}
\begin{figure}
	\centering
	\includegraphics{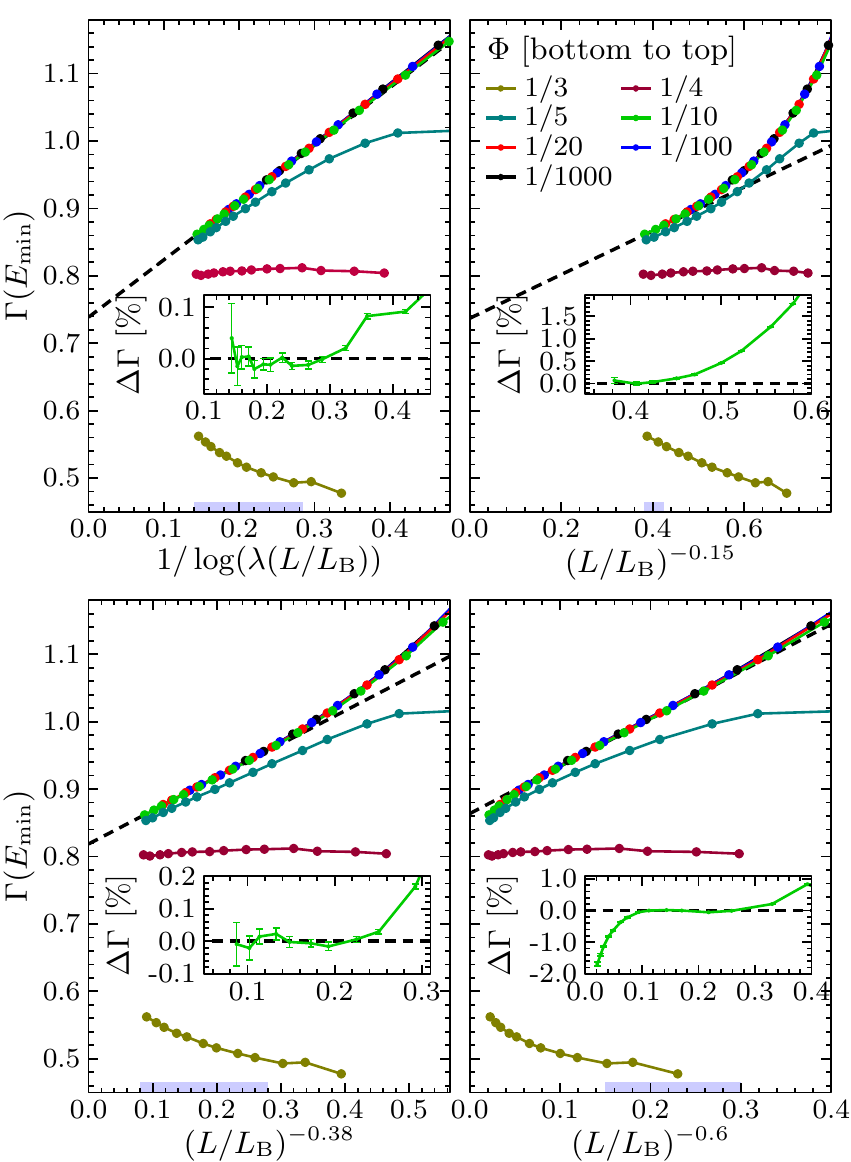}
	\caption{System-size dependence of $\Gamma$ at $E_{\text{min}}$ for several $\Phi$ under assumption of logarithmic scaling corrections (upper left panel) with $\lambda=1.68$ and power-law corrections (upper right and lower panels) with $y=0.15$, $0.38$, or $0.6$. The associated fits (black dashed lines) are based on subsets of the data ($\Phi\leq1/10$ with the $E$ range marked by the blue bars at the bottom of the panels). The insets show the deviations $\Delta\Gamma$ of the $\Phi=1/10$ data points from the fit function. Solid lines are guides to the eye only.}
	\label{fig:critical_behavior-ext}
\end{figure}
The IQH transition is known to feature strong corrections to scaling. In the literature \cite{SleO09S,NudKS15S,ObuGE12S,GruKN17S,SLEO12S,AmaMS11S}, the nature of the finite-size corrections is discussed controversially, and quantitative results vary strongly. Figure\ \ref{fig:critical_behavior-ext} addresses this question based on our data for the dimensionless Lyapunov exponent $\Gamma(E_\mathrm{min},\Phi,L)$. It compares logarithmic scaling corrections of the type $1/\log(\lambda(L/L_\mathrm{B}))$ 
and power-law corrections, $(L/L_\mathrm{B})^{-y}$, under the assumptions of $y=0.15$, $0.38$, and $0.60$, receptively. The data are plotted in such a way that the expected behaviors correspond to straight lines. In all cases, the data for $\Phi\lesssim1/10$ fall onto a common master curve that describes the universal regime where neither LL coupling nor the nonzero LL width play a role. We focus on this universal curve and consider our most precise data, $\Phi=1/10$, for a detailed analysis. For power-law corrections, the value $y=0.6$ leads to an approximately straight line for $8 L_\mathrm{B}\lesssim L\lesssim 44 L_\mathrm{B}$ but a downturn is clearly visible for larger $L/L_\mathrm{B}$. This demonstrates that $y\gtrsim 0.6$ is only an effective value holding over a limited size range. This downturn is not observed for $y=0.38$ and the power-law behavior with $L\gtrsim44L_\mathrm{B}$ yields $\Gamma_\mathrm{c}=0.818(1)$ based on a least-squares fit with $N_\mathrm{C}=2$ fit parameters and $N_\mathrm{P}=8$ data points and reduced error sum $\tilde{\chi}=0.9$ (see definition in analogy to Eq.\ \ref{eq:FSS_chi2}). The graph with $y=0.15$ does not give a straight-line behavior within our size range. 
Fits using a free $y$ exponent ($N_\mathrm{C}=3$) are robust under varying the range of considered system sizes, and we summarize the estimates as $\Gamma_\mathrm{c}=0.816(4)$ and $y=0.36(3)$ ($\tilde{\chi}\approx 0.8$ and $N_\mathrm{P}=6$ or $8$). For small values of $y$, it is clear that higher correction orders may have a large influence. However, including the next higher correction term, $(L/L_\mathrm{B})^{-2y}$, does not lead to significantly improved fits for a broader range of system sizes and estimates for $\Gamma_\mathrm{c}$ and $y$ are not substantially changed. 

For logarithmic scaling corrections, the scale factor $\lambda$ plays a similar role as $y$ for power-law corrections. Since its value is unknown, we use $\lambda$ as a free parameter in the fit function. We obtain fits of high quality for a broad range, $L\gtrsim 20L_\mathrm{B}$ ($N_\mathrm{P}=10$), of system sizes for $\Phi=1/10$, yielding $\lambda=1.68(4)$ and $\Gamma_\mathrm{c}=0.739(2)$ with $\tilde{\chi}=0.75$. Restricting the fits to large sizes only leads to the same $\Gamma_\mathrm{c}$ and $\lambda$ (within their error bars).

Based on the available data, we thus cannot qualitatively discriminate between power-law corrections with $y\approx 0.38$ and logarithmic scaling corrections. The estimates for $\Gamma_\mathrm{c}$ substantially depend on the type of irrelevant corrections.

\section{Details of the Finite-size scaling in the vicinity of the transition}

\begin{figure}
	\centering
	\includegraphics{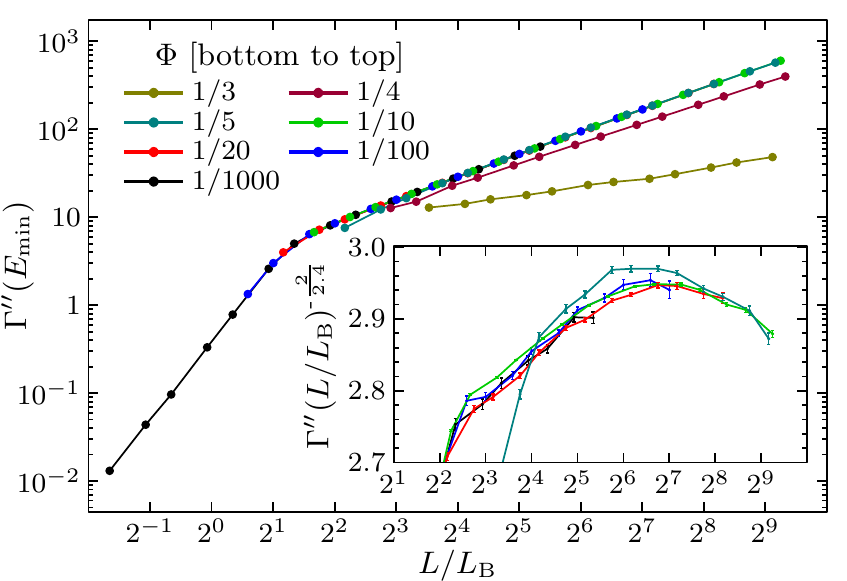}
	\caption{System-size dependence of the curvature $\Gamma''$ at $E_{\text{min}}$ for several $\Phi$. $\Gamma''$ data are scaled by a constant but $\Phi$-dependent parameter. Solid lines are guides to the eye only. Inset shows the system-size dependence of the curvature $\Gamma''$ reduced by power-law behavior with the exponent $2/\nu^\ast$ and assumed $\nu^\ast=2.4$.}
	\label{fig:curvature_scaling}
\end{figure}

To determine the localization length exponent $\nu$, we consider the system-size dependence of the curvature $\Gamma''(L)=\partial^2\Gamma/\partial^2 E\sim L^{2/\nu}$ at the minimum position $E_\mathrm{min}$ (with $E_\mathrm{min}\rightarrow E_\mathrm{c}$ for $L\rightarrow\infty$). Figure\ \ref{fig:curvature_scaling} shows the behavior for all fluxes $\Phi$. Similar to $\Gamma(E_\mathrm{c})$, the data of $\Gamma''$ follow a common line for $\Phi\lesssim1/10$. While the data for $L\lesssim 4L_\mathrm{B}$ are outside of the asymptotic scaling regime, the points for $L>4L_\mathrm{B}$ show only slight deviations from a straight power-law behavior. The inset of Fig.~\ref{fig:curvature_scaling} resolves them, by scaling $\Gamma''$ with $L^{2/2.4}$. This would remove the leading $L$ dependence corresponding to a localization length exponent $\nu=2.4$, observed in experiment and for the geometrically disordered Chalker-Coddington (CC) model. In other words, if $\nu=2.4$, the data in the inset should approach a horizontal line for $L/L_\mathrm{B}\rightarrow\infty$. While the scaled curvature data for $32 L_\mathrm{B} \lesssim L\lesssim 128 L_\mathrm{B}$ may be interpreted in such a way, the pronounced downturn of the data for $L> 128 L_\mathrm{B}$ shows that the asymptotic $\nu$ exceeds $2.4$.
The data for $\Phi=1/10$ and $L\geq 64$ ($N_\mathrm{P}=8$) can be accurately fitted with $\Gamma''(L)=a (1-bL^{-y}) L^{2/\nu}$. If we fix $\nu=2.55$ or $\nu=2.60$ ($N_\mathrm{C}=3$), we observe $y=0.32(2)$ with $\tilde{\chi}=1.04$ or $y=0.21(2)$ with $\tilde{\chi}=1.02$), respectively. For flux $\Phi=1/5$, $\Gamma''(L/L_\mathrm{B})$ deviates from the universal line only slightly. A fit based on $L\geq 24$ ($N_\mathrm{P}=10$) yields $\nu=2.55(8)$ and $y=0.4(3)$ with $\tilde{\chi}=1.02$ and $N_\mathrm{C}=4$. For $\Phi>1/5$, $\Gamma''(L/L_\mathrm{B})$ deviates from the universal curve, giving a nonuniversal $\nu$ that is higher than that for the universal case. This analysis of the curvature data provides valuable insights, but fits with two floating exponents lead to imprecise numerical exponent values.      

In a more sophisticated approach, we combine the determination of the critical exponents with the analysis of $\Gamma_\mathrm{c}$. Here, we consider the joint energy and system-size dependence of the dimensionless Lyapunov exponent $\Gamma$ in the vicinity of the critical point $E_\mathrm{c}$ and use the compact scaling ansatz
\begin{eqnarray}
\Gamma(x_{\text{r}} L^{1/\nu}, x_{\text{i}} L^{-y})=\sum\limits_{i=0}^{n_{\text{i}}}\sum\limits_{j=0}^{n_{\text{r}}}\frac{a_{ij}x_{\text{i}}^{i}x_{\text{r}}^{2 j}}{i!(2j)!}L^{2 j /\nu-iy}\label{eq:fss}
\end{eqnarray}
with the relevant scaling field $x_{\text{r}} L^{1/\nu}$ and a single irrelevant scaling field $x_{\text{i}} L^{-y}$. Since the localized phases on both sides of the transition behave similarly, the expansion of $\Gamma$ in terms of the relevant scaling field has only even terms (similar to recent investigations of the CC model, see e.g. \cite{SleO09S,NudKS15S,ObuGE12S,GruKN17S}). However, in contrast to the CC model, our data away from $E_\mathrm{c}$ are not perfectly symmetric in $E$ with respect to $E_\mathrm{c}$. Therefore, a general expansion of the relevant and irrelevant scaling variables
\begin{eqnarray}
x_{\text{r}}=e+\sum\limits_{k=2}^{m_{\text{r}}}\frac{b_{k}}{k!}e^k\quad\text{and}\quad x_{\text{i}}=1+\sum\limits_{l=1}^{m_{\text{i}}}\frac{c_{l}}{l!}e^l \label{eq:FSS_variables}
\end{eqnarray}
as functions of $e=E/E_\mathrm{c}-1$ needs to include both even and odd terms in $e$. The $N_\mathrm{C}=(n_{\text{i}}+1)(n_{\text{r}}+1)+m_{\text{i}}+m_{\text{r}}+\Theta(n_{\text{i}}) +1)$ coefficients of the expansion (\ref{eq:fss}) are found via a single compact fit by the method of least squares. Thereby, the reduced error sum
\begin{eqnarray}
\tilde{\chi}^2=\frac{1}{N_\mathrm{F}}\sum\limits_{i}^{N_\mathrm{P}} \frac{\left(\Gamma_i-\Gamma(L_i, E_i)\right)^2}{\sigma_i^2}\label{eq:FSS_chi2}
\end{eqnarray}
between the scaling function $\Gamma$ and $N_\mathrm{P}$ data points $\Gamma_i$ with standard deviation $\sigma_i$, observed for $E_i$ and $L_i$, will be minimized. We consider the fit to be of good quality if $\tilde{\chi}^2\approx 1.0$ for $N_\mathrm{F}=N_\mathrm{P}-N_\mathrm{C}\gg1$. The uncertainties of the fit parameter represent the standard deviation of the fit parameters based on $1000$ synthetic datasets. Each dataset arises from a different realization of Gaussian random noise added to the original data points, $\Gamma_i$. The standard deviation of the noise is equal to the uncertainty $\sigma_i$ of the individual data point $i$. 

For $\Phi=1/10$, the $\Gamma$ data in vicinity of $E_\mathrm{c}=3.422151(3)$ consist of $50$ energy points for each $L$ (see Fig.\ 3 of the main text). Neglecting irrelevant corrections, $n_\mathrm{i}=m_\mathrm{i}=0$, an expansion with $n_\mathrm{r}=m_\mathrm{r}=3$ applied to the two largest systems, $L\geq512$, yields $\nu=2.595(12)$ and $\Gamma_\mathrm{c}=0.8669(3)$ with moderate accuracy $\tilde{\chi}=1.4$ ($N_\mathrm{P}=100$ and $N_\mathrm{C}=8$). This fit is clearly insufficient as we have seen above that corrections to scaling are strong. Including simple irrelevant corrections, $n_\mathrm{i}=1$ and $m_\mathrm{i}=0$, and using the data with $L\geq128$ ($N_\mathrm{P}=300$) yields $\Gamma_\mathrm{c}=0.814(6)$, $\nu=2.594(17)$, and $y=0.357(26)$ with $\tilde{\chi}=1.02$ ($N_\mathrm{C}=13$), presenting our most reliable fit. While variation of $n_\mathrm{r}$ and $m_\mathrm{r}$ generally leads to results that agree within their error bars, a lower expansion, $m_\mathrm{r}=2$, of the relevant scaling field gives $\nu=2.558(10)$, based on a fit with slightly reduced quality ($\tilde{\chi}=1.04$ with $N_\mathrm{C}=12$) . We therefore consider $m_\mathrm{r}\geq 3$ for a suitable description of the relevant asymmetry of our data with respect to $E_\mathrm{c}$. 
The inclusion of smaller systems, e.g., $L\geq64$ ($N_\mathrm{P}=400$), leads to $\nu=2.558(18)$ and $y=0.368(16)$ with $\tilde{\chi}=1.07$. Higher irrelevant expansion orders, e.g., $n_\mathrm{i}=2$ or $m_\mathrm{i}>0$, lead to very small improvements of $\tilde{\chi}$ only. The resulting critical parameter values overlap with those quoted above, but their uncertainties increase. Taking our various fits into consideration, we estimate the critical parameters to $\Gamma_\mathrm{c}=0.815(8)$, $\nu=2.58(3)$, and $y=0.35(4)$. 

To describe our data based on logarithmic corrections, we replace the irrelevant field $x_\mathrm{i}L^{-y}$ in Eq. (\ref{eq:fss}) by $x_\mathrm{i}/(b+x_\mathrm{i}\log L)$. For $\Phi=1/10$, we observe the best fit for $L\geq128$ ($N_\mathrm{P}=300$) with the same expansion orders ($N_\mathrm{C}=13$) as above, yielding $\Gamma_\mathrm{c}=0.745(6)$ and $\nu=2.597(27)$ with $\tilde{\chi}=1.02$. Similar to fits with power-law corrections, the inclusion of smaller system sizes makes higher expansion orders necessary. For example, for $L\geq 24$ ( $N_\mathrm{P}=550$), we used with $n_\mathrm{i}=2$, $m_\mathrm{i}=1$, and $n_\mathrm{r}=m_\mathrm{r}=3$ $18$ parameters, yielding $\Gamma_\mathrm{c}=0.740(5)$ and $\nu=2.653(35)$ with $\tilde{\chi}=1.03$.

The description of our data based on logarithmic corrections gives reliable fits in the same limits as for power-law corrections. Hence, we cannot discriminate between the two correction types. Our estimate of  $\Gamma_\mathrm{c}$ depends on the correction type. However, the values of $\Gamma_\mathrm{c}$ (and $y$) based on the sophisticated scaling approach are consistent with the above graphical investigation of finite-size corrections at the critical point. Interestingly, our results for $\nu$ do not depend on the type of correction, giving us confidence in its value.      

\section{Transition in cases of non-negligible intrinsic LL width}
\begin{figure}
	\includegraphics{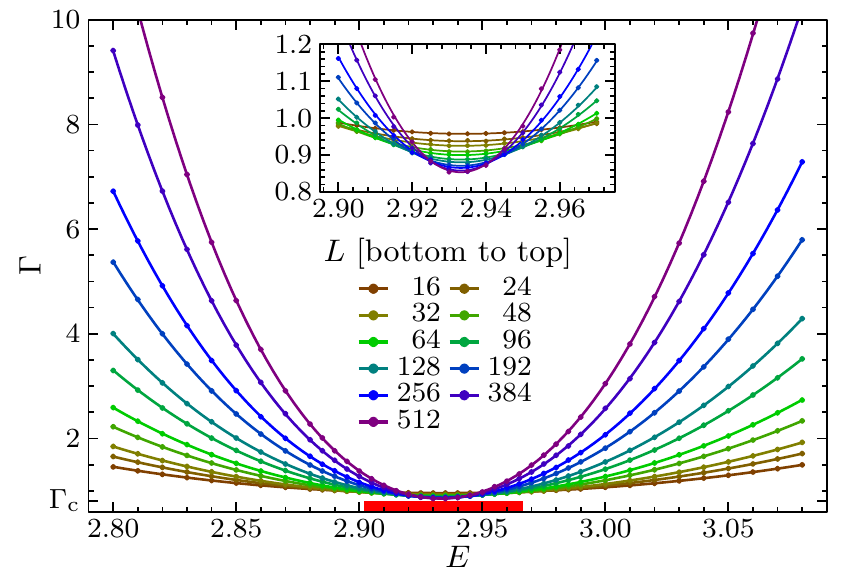}
	\includegraphics{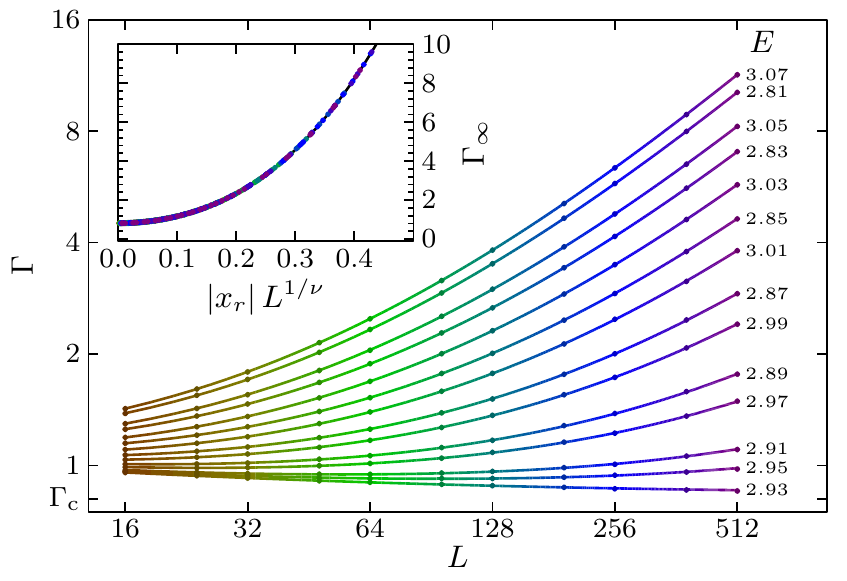}
	\caption{$\Gamma(E,L)$ for $\Phi=1/5$ as function of $E$ for several $L$ (upper panel) and as function of $L$ for several $E$ (lower panel). The statistical errors are well below the symbol size. The solid lines result from a fit, see text for details. The red bar at the bottom shows the energy range of the corresponding LL in the absence of disorder (see Fig. 2 of the main text). The insets show a magnification of the immediate vicinity of the transition (upper panel) and the asymptotic scaling function (lower panel).
	}
	\label{fig:transition_1o5}
\end{figure}

\begin{figure}
	\includegraphics{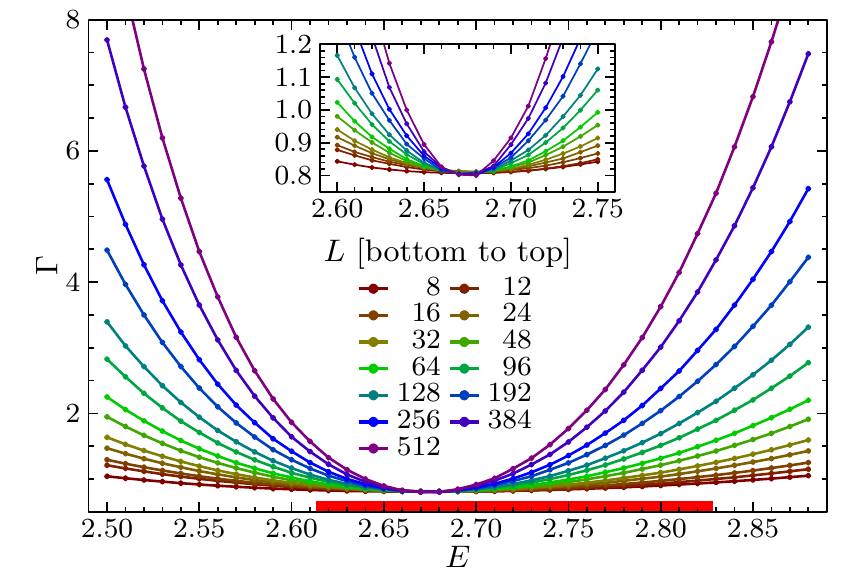}
	\includegraphics{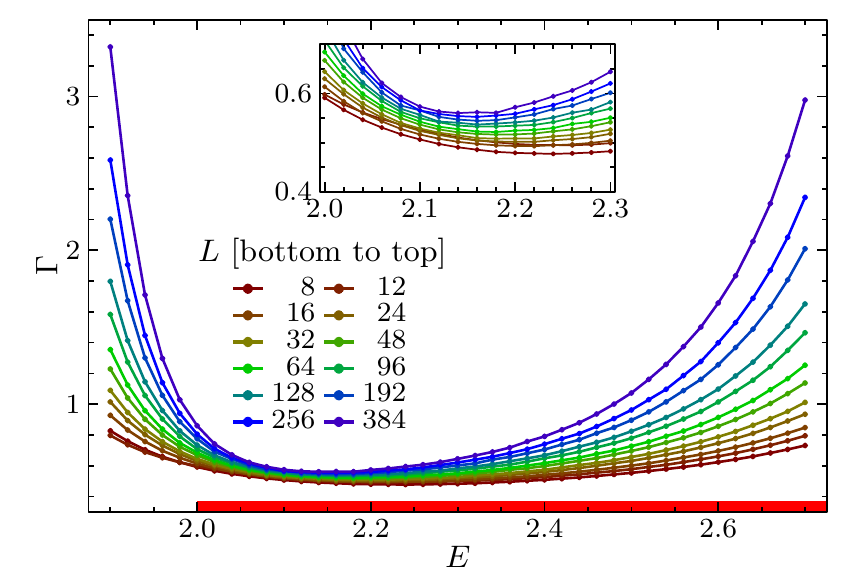}	
	\caption{$\Gamma(E,L)$ in vicinity of the IQH-like transition for $\Phi=1/4$ (upper panel) and $1/3$ (lower panel) as function of $E$ for several $L$. The statistical errors are well below the symbol size. Solid lines are guide to the eye only. The red bars at the bottom show the energy range of the corresponding LL in the absence of disorder (see Fig. 2 of the main text). Insets show a magnification of the immediate vicinity of $E_\mathrm{c}$. 
	}
	\label{fig:transition_1o3_1o4}
\end{figure}

For $\Phi=1/5$, $1/4$, and $1/3$, the disorder strength $W=2$ leads to qualitatively similar LBs (see Fig.\ \ref{fig:square_dos_phi}). There is a gap between the lowest and the first LB, so inter-LB scattering may still be negligible. Figure~\ref{fig:transition_1o5} shows $\Gamma(E,\Phi,L)$ in the vicinity of the critical point for $\Phi=1/5$, and Fig.~\ref{fig:transition_1o3_1o4} does the same for $\Phi=1/4$ and $1/3$. With increasing $\Phi$ (i.e., increasing $\delta$), the asymmetry of $\Gamma(E)$ with respect to $E_\mathrm{c}$ increases, and the system-size dependence of $\Gamma(E_\mathrm{c},L)$ changes sign.

For $\Phi=1/5$, the above graphical investigations for $\Gamma(E_\mathrm{c})$ and $\Gamma''(E_\mathrm{c})$ indicate behaviors similar to the universal data, $\Phi\lesssim 1/10$. However, caused by the non-negligible intrinsic LL width, the appropriate compact scaling ansatz for the data in vicinity of the transition is not clear. To identify the appropriate form of the relevant scaling field, the lower panel of Fig.\ \ref{fig:transition_1o5} visualizes $\Gamma(E,L)$ as function of $L$ for several $E$ in vicinity of the transition. It shows that one can find pairs ${E_1,\ E_2}$ of energy values with $E_1<E_\mathrm{c}<E_2$ so that $\Gamma(E_1,L)=\Gamma(E_2,L)$ (for sufficiently large $L$). Therefore, the expansion\ (\ref{eq:fss}) of $\Gamma$ in even powers of the relevant scaling field is still suitable. Moreover, we use the same expansion (\ref{eq:FSS_variables}) of the relevant scaling variable to describe the pronounced asymmetry in the dependence of $E$ with respect to $E_\mathrm{c}$.

The position of the IQH transition is less affected by expansion of the fit functions and we determine $E_\mathrm{c}=2.93316(3)$. For $L\geq128$ ($N_\mathrm{P}=165$) we use with $n_\mathrm{i}=1$, $m_\mathrm{i}=0$, and $n_\mathrm{r}=m_\mathrm{r}=3$ ($N_\mathrm{C}=13$), yielding $\Gamma_\mathrm{c}=0.805(27)$, $\nu=2.561(31)$, and $y=0.32(11)$ with $\tilde{\chi}=0.96$. In comparison to the universal data, we observe suitable fits for a broader range of system sizes. For example, with $n_\mathrm{i}=1$, $m_\mathrm{i}=1$, and $n_\mathrm{r}=m_\mathrm{r}=3$ ($N_\mathrm{C}=14$) for $L\geq16$ ($N_\mathrm{P}=363$) give $\Gamma_\mathrm{c}=0.815(2)$, $\nu=2.547(3)$, and $y=0.360(6)$ with $\tilde{\chi}=1.00$. Taking into account the robustness of the fit against changes in size range and expansion orders, we estimate the critical parameters to be $\Gamma_\mathrm{c}=0.810(10)$, $\nu=2.55(4)$, and $y=0.33(5)$. This means they overlap with the results for the universal case.

For logarithmic corrections to scaling, we obtain $\Gamma_\mathrm{c}=0.731(16)$ and $\nu=2.570(37)$ with $\tilde{\chi}=0.98$ using $L\geq128$, and $\Gamma_\mathrm{c}=0.733(8)$ and $\nu=2.548(10)$ with $\tilde{\chi}=0.95$ using $L\geq16$ with the same expansion orders as for power-law corrections.

For $\Phi=1/4$, $\Gamma(E_\mathrm{c},L)$ appears to be almost independent of $L$. However, finite-size effects are still present, requiring asymmetric terms $m_\mathrm{i}>0$. Assuming corrections to scaling with a single irrelevant field, our fits based on different data ranges yield consistent parameters but are of moderate quality ($\tilde{\chi}\approx 1.2$) only. The result of the scaling analysis can be summarized as $\nu=2.70(3)$, $y=0.30(7)$, and $\Gamma_\mathrm{c}=0.795(8)$.

For $\Phi=1/3$, corrections to scaling are much more pronounced. The scaling behavior is strongly asymmetric with respect to $E_\mathrm{c}$, and the minimum position of $\Gamma(E,L)$ as function of $E$ has a significant dependence on $L$. We therefore use only larger systems, $L\geq 64$, and consider a sophisticated expansion of the scaling variables, i.e., $m_\mathrm{i}$ up to 3 and $m_\mathrm{r}$ up to 6. However, our best fits are of reduced quality ($\tilde{\chi}\approx 1.2$). If taken at face value, they suggest a significantly increased relevant exponent $\nu=4.0(5)$ with $y=0.13(3)$, and $\Gamma_\mathrm{c}=0.70(5)$.

In general, the analysis of transitions with non-negligible intrinsic LL is challenging. The scaling behavior seems to be affected by several kinds of finite-size effects, so that a scaling approach with a single irrelevant field may be insufficient. Based on our data we cannot determine whether the results for $\Phi>1/5$ correspond to effective behavior only, with the asymptotic behavior reached only for much larger system sizes or whether the asymptotic exponents for large $\Phi$ are truly nonuniversal, corresponding to smooth crossover to the case $\Phi=1/2$ where no IQH transition occurs because the two LLs touch in the disorder-free case. 


\begin{thebibliography}{50}%
		\makeatletter
		\providecommand \@ifxundefined [1]{%
			\@ifx{#1\undefined}
		}%
		\providecommand \@ifnum [1]{%
			\ifnum #1\expandafter \@firstoftwo
			\else \expandafter \@secondoftwo
			\fi
		}%
		\providecommand \@ifx [1]{%
			\ifx #1\expandafter \@firstoftwo
			\else \expandafter \@secondoftwo
			\fi
		}%
		\providecommand \natexlab [1]{#1}%
		\providecommand \enquote  [1]{``#1''}%
		\providecommand \bibnamefont  [1]{#1}%
		\providecommand \bibfnamefont [1]{#1}%
		\providecommand \citenamefont [1]{#1}%
		\providecommand \href@noop [0]{\@secondoftwo}%
		\providecommand \href [0]{\begingroup \@sanitize@url \@href}%
		\providecommand \@href[1]{\@@startlink{#1}\@@href}%
		\providecommand \@@href[1]{\endgroup#1\@@endlink}%
		\providecommand \@sanitize@url [0]{\catcode `\\12\catcode `\$12\catcode
			`\&12\catcode `\#12\catcode `\^12\catcode `\_12\catcode `\%12\relax}%
		\providecommand \@@startlink[1]{}%
		\providecommand \@@endlink[0]{}%
		\providecommand \url  [0]{\begingroup\@sanitize@url \@url }%
		\providecommand \@url [1]{\endgroup\@href {#1}{\urlprefix }}%
		\providecommand \urlprefix  [0]{URL }%
		\providecommand \Eprint [0]{\href }%
		\providecommand \doibase [0]{http://dx.doi.org/}%
		\providecommand \selectlanguage [0]{\@gobble}%
		\providecommand \bibinfo  [0]{\@secondoftwo}%
		\providecommand \bibfield  [0]{\@secondoftwo}%
		\providecommand \translation [1]{[#1]}%
		\providecommand \BibitemOpen [0]{}%
		\providecommand \bibitemStop [0]{}%
		\providecommand \bibitemNoStop [0]{.\EOS\space}%
		\providecommand \EOS [0]{\spacefactor3000\relax}%
		\providecommand \BibitemShut  [1]{\csname bibitem#1\endcsname}%
		\let\auto@bib@innerbib\@empty
		%</preamble>
		\bibitem [{\citenamefont {Klitzing}\ \emph {et~al.}(1980)\citenamefont
			{Klitzing}, \citenamefont {Dorda},\ and\ \citenamefont {Pepper}}]{KliDP80}%
		\BibitemOpen
		\bibfield  {author} {\bibinfo {author} {\bibfnamefont {K.~v.}\ \bibnamefont
				{Klitzing}}, \bibinfo {author} {\bibfnamefont {G.}~\bibnamefont {Dorda}}, \
			and\ \bibinfo {author} {\bibfnamefont {M.}~\bibnamefont {Pepper}},\ }\href
		{\doibase 10.1103/PhysRevLett.45.494} {\bibfield  {journal} {\bibinfo
				{journal} {Phys. Rev. Lett.}\ }\textbf {\bibinfo {volume} {45}},\ \bibinfo
			{pages} {494} (\bibinfo {year} {1980})}\BibitemShut {NoStop}%
		\bibitem [{\citenamefont {Anderson}(1958)}]{And58}%
		\BibitemOpen
		\bibfield  {author} {\bibinfo {author} {\bibfnamefont {P.~W.}\ \bibnamefont
				{Anderson}},\ }\href@noop {} {\bibfield  {journal} {\bibinfo  {journal}
				{Phys. Rev.}\ }\textbf {\bibinfo {volume} {109}},\ \bibinfo {pages} {1492}
			(\bibinfo {year} {1958})}\BibitemShut {NoStop}%
		\bibitem [{\citenamefont {Evers}\ and\ \citenamefont {Mirlin}(2008)}]{EveM08}%
		\BibitemOpen
		\bibfield  {author} {\bibinfo {author} {\bibfnamefont {F.}~\bibnamefont
				{Evers}}\ and\ \bibinfo {author} {\bibfnamefont {A.~D.}\ \bibnamefont
				{Mirlin}},\ }\href {\doibase 10.1103/RevModPhys.80.1355} {\bibfield
			{journal} {\bibinfo  {journal} {Rev. Mod. Phys.}\ }\textbf {\bibinfo {volume}
				{80}},\ \bibinfo {pages} {1355} (\bibinfo {year} {2008})}\BibitemShut
		{NoStop}%
		\bibitem [{\citenamefont {Li}\ \emph {et~al.}(2005)\citenamefont {Li},
			\citenamefont {Cs\'athy}, \citenamefont {Tsui}, \citenamefont {Pfeiffer},\
			and\ \citenamefont {West}}]{LiCT05}%
		\BibitemOpen
		\bibfield  {author} {\bibinfo {author} {\bibfnamefont {W.}~\bibnamefont
				{Li}}, \bibinfo {author} {\bibfnamefont {G.~A.}\ \bibnamefont {Cs\'athy}},
			\bibinfo {author} {\bibfnamefont {D.~C.}\ \bibnamefont {Tsui}}, \bibinfo
			{author} {\bibfnamefont {L.~N.}\ \bibnamefont {Pfeiffer}}, \ and\ \bibinfo
			{author} {\bibfnamefont {K.~W.}\ \bibnamefont {West}},\ }\href {\doibase
			10.1103/PhysRevLett.94.206807} {\bibfield  {journal} {\bibinfo  {journal}
				{Phys. Rev. Lett.}\ }\textbf {\bibinfo {volume} {94}},\ \bibinfo {pages}
			{206807} (\bibinfo {year} {2005})}\BibitemShut {NoStop}%
		\bibitem [{\citenamefont {Li}\ \emph {et~al.}(2009{\natexlab{a}})\citenamefont
			{Li}, \citenamefont {Vicente}, \citenamefont {Xia}, \citenamefont {Pan},
			\citenamefont {Tsui}, \citenamefont {Pfeiffer},\ and\ \citenamefont
			{West}}]{LiVX09}%
		\BibitemOpen
		\bibfield  {author} {\bibinfo {author} {\bibfnamefont {W.}~\bibnamefont
				{Li}}, \bibinfo {author} {\bibfnamefont {C.~L.}\ \bibnamefont {Vicente}},
			\bibinfo {author} {\bibfnamefont {J.~S.}\ \bibnamefont {Xia}}, \bibinfo
			{author} {\bibfnamefont {W.}~\bibnamefont {Pan}}, \bibinfo {author}
			{\bibfnamefont {D.~C.}\ \bibnamefont {Tsui}}, \bibinfo {author}
			{\bibfnamefont {L.~N.}\ \bibnamefont {Pfeiffer}}, \ and\ \bibinfo {author}
			{\bibfnamefont {K.~W.}\ \bibnamefont {West}},\ }\href {\doibase
			10.1103/PhysRevLett.102.216801} {\bibfield  {journal} {\bibinfo  {journal}
				{Phys. Rev. Lett.}\ }\textbf {\bibinfo {volume} {102}},\ \bibinfo {pages}
			{216801} (\bibinfo {year} {2009}{\natexlab{a}})}\BibitemShut {NoStop}%
		\bibitem [{\citenamefont {Giesbers}\ \emph {et~al.}(2009)\citenamefont
			{Giesbers}, \citenamefont {Zeitler}, \citenamefont {Ponomarenko},
			\citenamefont {Yang}, \citenamefont {Novoselov}, \citenamefont {Geim},\ and\
			\citenamefont {Maan}}]{GieZP09}%
		\BibitemOpen
		\bibfield  {author} {\bibinfo {author} {\bibfnamefont {A.~J.~M.}\
				\bibnamefont {Giesbers}}, \bibinfo {author} {\bibfnamefont {U.}~\bibnamefont
				{Zeitler}}, \bibinfo {author} {\bibfnamefont {L.~A.}\ \bibnamefont
				{Ponomarenko}}, \bibinfo {author} {\bibfnamefont {R.}~\bibnamefont {Yang}},
			\bibinfo {author} {\bibfnamefont {K.~S.}\ \bibnamefont {Novoselov}}, \bibinfo
			{author} {\bibfnamefont {A.~K.}\ \bibnamefont {Geim}}, \ and\ \bibinfo
			{author} {\bibfnamefont {J.~C.}\ \bibnamefont {Maan}},\ }\href {\doibase
			10.1103/PhysRevB.80.241411} {\bibfield  {journal} {\bibinfo  {journal} {Phys.
					Rev. B}\ }\textbf {\bibinfo {volume} {80}},\ \bibinfo {pages} {241411}
			(\bibinfo {year} {2009})}\BibitemShut {NoStop}%
		\bibitem [{\citenamefont {Huckestein}\ and\ \citenamefont
			{Kramer}(1990)}]{HucK90}%
		\BibitemOpen
		\bibfield  {author} {\bibinfo {author} {\bibfnamefont {B.}~\bibnamefont
				{Huckestein}}\ and\ \bibinfo {author} {\bibfnamefont {B.}~\bibnamefont
				{Kramer}},\ }\href@noop {} {\bibfield  {journal} {\bibinfo  {journal} {Phys.
					Rev. Lett.}\ }\textbf {\bibinfo {volume} {64}},\ \bibinfo {pages} {1437}
			(\bibinfo {year} {1990})}\BibitemShut {NoStop}%
		\bibitem [{\citenamefont {Huckestein}(1992)}]{Huc92}%
		\BibitemOpen
		\bibfield  {author} {\bibinfo {author} {\bibfnamefont {B.}~\bibnamefont
				{Huckestein}},\ }\href@noop {} {\bibfield  {journal} {\bibinfo  {journal}
				{Europhys. Lett.}\ }\textbf {\bibinfo {volume} {20}},\ \bibinfo {pages} {451}
			(\bibinfo {year} {1992})}\BibitemShut {NoStop}%
		\bibitem [{\citenamefont {Huckestein}(1995)}]{Huc95}%
		\BibitemOpen
		\bibfield  {author} {\bibinfo {author} {\bibfnamefont {B.}~\bibnamefont
				{Huckestein}},\ }\href@noop {} {\bibfield  {journal} {\bibinfo  {journal}
				{Rev. Mod. Phys.}\ }\textbf {\bibinfo {volume} {67}},\ \bibinfo {pages} {357}
			(\bibinfo {year} {1995})}\BibitemShut {NoStop}%
		\bibitem [{\citenamefont {Lee}\ and\ \citenamefont {Wang}(1996)}]{LeeW96}%
		\BibitemOpen
		\bibfield  {author} {\bibinfo {author} {\bibfnamefont {D.-H.}\ \bibnamefont
				{Lee}}\ and\ \bibinfo {author} {\bibfnamefont {Z.}~\bibnamefont {Wang}},\
		}\href@noop {} {\bibfield  {journal} {\bibinfo  {journal} {Phys. Rev. Lett.}\
			}\textbf {\bibinfo {volume} {76}},\ \bibinfo {pages} {4014} (\bibinfo {year}
			{1996})}\BibitemShut {NoStop}%
		\bibitem [{\citenamefont {Huo}\ and\ \citenamefont {Bhatt}(1992)}]{HuoB92}%
		\BibitemOpen
		\bibfield  {author} {\bibinfo {author} {\bibfnamefont {Y.}~\bibnamefont
				{Huo}}\ and\ \bibinfo {author} {\bibfnamefont {R.~N.}\ \bibnamefont
				{Bhatt}},\ }\href@noop {} {\bibfield  {journal} {\bibinfo  {journal} {Phys.
					Rev. Lett.}\ }\textbf {\bibinfo {volume} {68}},\ \bibinfo {pages} {1375}
			(\bibinfo {year} {1992})}\BibitemShut {NoStop}%
		\bibitem [{\citenamefont {Cain}\ \emph {et~al.}(2003)\citenamefont {Cain},
			\citenamefont {{R\"{o}mer}},\ and\ \citenamefont {Raikh}}]{CaiRR03}%
		\BibitemOpen
		\bibfield  {author} {\bibinfo {author} {\bibfnamefont {P.}~\bibnamefont
				{Cain}}, \bibinfo {author} {\bibfnamefont {R.~A.}\ \bibnamefont
				{{R\"{o}mer}}}, \ and\ \bibinfo {author} {\bibfnamefont {M.~E.}\ \bibnamefont
				{Raikh}},\ }\href@noop {} {\bibfield  {journal} {\bibinfo  {journal} {Phys.
					Rev. B}\ }\textbf {\bibinfo {volume} {67}},\ \bibinfo {pages} {075307}
			(\bibinfo {year} {2003})}\BibitemShut {NoStop}%
		\bibitem [{\citenamefont {Mkhitaryan}\ and\ \citenamefont
			{Raikh}(2009)}]{MkhR2009}%
		\BibitemOpen
		\bibfield  {author} {\bibinfo {author} {\bibfnamefont {V.~V.}\ \bibnamefont
				{Mkhitaryan}}\ and\ \bibinfo {author} {\bibfnamefont {M.~E.}\ \bibnamefont
				{Raikh}},\ }\href {\doibase 10.1103/PhysRevB.79.125401} {\bibfield  {journal}
			{\bibinfo  {journal} {Phys. Rev. B}\ }\textbf {\bibinfo {volume} {79}},\
			\bibinfo {pages} {125401} (\bibinfo {year} {2009})}\BibitemShut {NoStop}%
		\bibitem [{\citenamefont {Chalker}\ and\ \citenamefont
			{Coddington}(1988)}]{ChaC88}%
		\BibitemOpen
		\bibfield  {author} {\bibinfo {author} {\bibfnamefont {J.~T.}\ \bibnamefont
				{Chalker}}\ and\ \bibinfo {author} {\bibfnamefont {P.~D.}\ \bibnamefont
				{Coddington}},\ }\href@noop {} {\bibfield  {journal} {\bibinfo  {journal} {J.
					Phys.: Condens. Matter}\ }\textbf {\bibinfo {volume} {21}},\ \bibinfo {pages}
			{2665} (\bibinfo {year} {1988})}\BibitemShut {NoStop}%
		\bibitem [{\citenamefont {Slevin}\ and\ \citenamefont
			{Ohtsuki}(2009)}]{SleO09}%
		\BibitemOpen
		\bibfield  {author} {\bibinfo {author} {\bibfnamefont {K.}~\bibnamefont
				{Slevin}}\ and\ \bibinfo {author} {\bibfnamefont {T.}~\bibnamefont
				{Ohtsuki}},\ }\href {\doibase 10.1103/PhysRevB.80.041304} {\bibfield
			{journal} {\bibinfo  {journal} {Phys. Rev. B}\ }\textbf {\bibinfo {volume}
				{80}},\ \bibinfo {pages} {041304} (\bibinfo {year} {2009})}\BibitemShut
		{NoStop}%
		\bibitem [{\citenamefont {Li}\ \emph {et~al.}(2009{\natexlab{b}})\citenamefont
			{Li}, \citenamefont {Vicente}, \citenamefont {Xia}, \citenamefont {Pan},
			\citenamefont {Tsui}, \citenamefont {Pfeiffer},\ and\ \citenamefont
			{West}}]{LiVX2009}%
		\BibitemOpen
		\bibfield  {author} {\bibinfo {author} {\bibfnamefont {W.}~\bibnamefont
				{Li}}, \bibinfo {author} {\bibfnamefont {C.~L.}\ \bibnamefont {Vicente}},
			\bibinfo {author} {\bibfnamefont {J.~S.}\ \bibnamefont {Xia}}, \bibinfo
			{author} {\bibfnamefont {W.}~\bibnamefont {Pan}}, \bibinfo {author}
			{\bibfnamefont {D.~C.}\ \bibnamefont {Tsui}}, \bibinfo {author}
			{\bibfnamefont {L.~N.}\ \bibnamefont {Pfeiffer}}, \ and\ \bibinfo {author}
			{\bibfnamefont {K.~W.}\ \bibnamefont {West}},\ }\href {\doibase
			10.1103/PhysRevLett.102.216801} {\bibfield  {journal} {\bibinfo  {journal}
				{Phys. Rev. Lett.}\ }\textbf {\bibinfo {volume} {102}},\ \bibinfo {pages}
			{216801} (\bibinfo {year} {2009}{\natexlab{b}})}\BibitemShut {NoStop}%
		\bibitem [{\citenamefont {Slevin}\ and\ \citenamefont
			{Ohtsuki}(2012)}]{SleO12}%
		\BibitemOpen
		\bibfield  {author} {\bibinfo {author} {\bibfnamefont {K.}~\bibnamefont
				{Slevin}}\ and\ \bibinfo {author} {\bibfnamefont {T.}~\bibnamefont
				{Ohtsuki}},\ }\href {\doibase 10.1142/S2010194512006162} {
			\bibfield {journal} {\bibinfo {journal} {Int. J. Phys. Conf. Ser.}\ }
			\textbf
			{\bibinfo {volume} {11}},\ \bibinfo {pages} {60} (\bibinfo {year}
			{2012})}\BibitemShut {NoStop}%	
		\bibitem [{\citenamefont {Obuse}\ \emph {et~al.}(2010)\citenamefont {Obuse},
			\citenamefont {Subramaniam}, \citenamefont {Furusaki}, \citenamefont
			{Gruzberg},\ and\ \citenamefont {Ludwig}}]{ObuSF10}%
		\BibitemOpen
		\bibfield  {author} {\bibinfo {author} {\bibfnamefont {H.}~\bibnamefont
				{Obuse}}, \bibinfo {author} {\bibfnamefont {A.~R.}\ \bibnamefont
				{Subramaniam}}, \bibinfo {author} {\bibfnamefont {A.}~\bibnamefont
				{Furusaki}}, \bibinfo {author} {\bibfnamefont {I.~A.}\ \bibnamefont
				{Gruzberg}}, \ and\ \bibinfo {author} {\bibfnamefont {A.~W.~W.}\ \bibnamefont
				{Ludwig}},\ }\href {\doibase 10.1103/PhysRevB.82.035309} {\bibfield
			{journal} {\bibinfo  {journal} {Phys. Rev. B}\ }\textbf {\bibinfo {volume}
				{82}},\ \bibinfo {pages} {035309} (\bibinfo {year} {2010})}\BibitemShut
		{NoStop}%
		\bibitem [{\citenamefont {Amado}\ \emph {et~al.}(2011)\citenamefont {Amado},
			\citenamefont {Malyshev}, \citenamefont {Sedrakyan},\ and\ \citenamefont
			{Dominguez-Adame}}]{AmaMS11}%
		\BibitemOpen
		\bibfield  {author} {\bibinfo {author} {\bibfnamefont {M.}~\bibnamefont
				{Amado}}, \bibinfo {author} {\bibfnamefont {A.~V.}\ \bibnamefont {Malyshev}},
			\bibinfo {author} {\bibfnamefont {A.}~\bibnamefont {Sedrakyan}}, \ and\
			\bibinfo {author} {\bibfnamefont {F.}~\bibnamefont {Dominguez-Adame}},\
		}\href {\doibase 10.1103/PhysRevLett.107.066402} {\bibfield  {journal}
			{\bibinfo  {journal} {Phys. Rev. Lett.}\ }\textbf {\bibinfo {volume} {107}},\
			\bibinfo {pages} {066402} (\bibinfo {year} {2011})}\BibitemShut {NoStop}%
		\bibitem [{\citenamefont {Fulga}\ \emph {et~al.}(2011)\citenamefont {Fulga},
			\citenamefont {Hassler}, \citenamefont {Akhmerov},\ and\ \citenamefont
			{Beenakker}}]{FulHA11}%
		\BibitemOpen
		\bibfield  {author} {\bibinfo {author} {\bibfnamefont {I.~C.}\ \bibnamefont
				{Fulga}}, \bibinfo {author} {\bibfnamefont {F.}~\bibnamefont {Hassler}},
			\bibinfo {author} {\bibfnamefont {A.~R.}\ \bibnamefont {Akhmerov}}, \ and\
			\bibinfo {author} {\bibfnamefont {C.~W.~J.}\ \bibnamefont {Beenakker}},\
		}\href {\doibase 10.1103/PhysRevB.84.245447} {\bibfield  {journal} {\bibinfo
				{journal} {Phys. Rev. B}\ }\textbf {\bibinfo {volume} {84}},\ \bibinfo
			{pages} {245447} (\bibinfo {year} {2011})}\BibitemShut {NoStop}%
		\bibitem [{\citenamefont {Obuse}\ \emph {et~al.}(2012)\citenamefont {Obuse},
			\citenamefont {Gruzberg},\ and\ \citenamefont {Evers}}]{ObuGE12}%
		\BibitemOpen
		\bibfield  {author} {\bibinfo {author} {\bibfnamefont {H.}~\bibnamefont
				{Obuse}}, \bibinfo {author} {\bibfnamefont {I.~A.}\ \bibnamefont {Gruzberg}},
			\ and\ \bibinfo {author} {\bibfnamefont {F.}~\bibnamefont {Evers}},\ }\href
		{\doibase 10.1103/PhysRevLett.109.206804} {\bibfield  {journal} {\bibinfo
				{journal} {Phys. Rev. Lett.}\ }\textbf {\bibinfo {volume} {109}},\ \bibinfo
			{pages} {206804} (\bibinfo {year} {2012})}\BibitemShut {NoStop}%
		\bibitem [{\citenamefont {Nuding}\ \emph {et~al.}(2015)\citenamefont {Nuding},
			\citenamefont {Kl\"umper},\ and\ \citenamefont {Sedrakyan}}]{NudKS15}%
		\BibitemOpen
		\bibfield  {author} {\bibinfo {author} {\bibfnamefont {W.}~\bibnamefont
				{Nuding}}, \bibinfo {author} {\bibfnamefont {A.}~\bibnamefont {Kl\"umper}}, \
			and\ \bibinfo {author} {\bibfnamefont {A.}~\bibnamefont {Sedrakyan}},\ }\href
		{\doibase 10.1103/PhysRevB.91.115107} {\bibfield  {journal} {\bibinfo
				{journal} {Phys. Rev. B}\ }\textbf {\bibinfo {volume} {91}},\ \bibinfo
			{pages} {115107} (\bibinfo {year} {2015})}\BibitemShut {NoStop}%
		\bibitem [{\citenamefont {Polyakov}\ and\ \citenamefont
			{Shklovskii}(1993)}]{PolS93}%
		\BibitemOpen
		\bibfield  {author} {\bibinfo {author} {\bibfnamefont {D.~G.}\ \bibnamefont
				{Polyakov}}\ and\ \bibinfo {author} {\bibfnamefont {B.~I.}\ \bibnamefont
				{Shklovskii}},\ }\href {\doibase 10.1103/PhysRevLett.70.3796} {\bibfield
			{journal} {\bibinfo  {journal} {Phys. Rev. Lett.}\ }\textbf {\bibinfo
				{volume} {70}},\ \bibinfo {pages} {3796} (\bibinfo {year}
			{1993})}\BibitemShut {NoStop}%
		\bibitem [{\citenamefont {Pruisken}\ and\ \citenamefont
			{Baranov}(1995)}]{PruB95}%
		\BibitemOpen
		\bibfield  {author} {\bibinfo {author} {\bibfnamefont {A.~M.~M.}\
				\bibnamefont {Pruisken}}\ and\ \bibinfo {author} {\bibfnamefont {M.~A.}\
				\bibnamefont {Baranov}},\ }\href
		{http://stacks.iop.org/0295-5075/31/i=9/a=007} {\bibfield  {journal}
			{\bibinfo  {journal} {Europhysics Letters}\ }\textbf {\bibinfo {volume}
				{31}},\ \bibinfo {pages} {543} (\bibinfo {year} {1995})}\BibitemShut
		{NoStop}%
		\bibitem [{\citenamefont {Wang}\ \emph {et~al.}(2000)\citenamefont {Wang},
			\citenamefont {Fisher}, \citenamefont {Girvin},\ and\ \citenamefont
			{Chalker}}]{ZiqFG00}%
		\BibitemOpen
		\bibfield  {author} {\bibinfo {author} {\bibfnamefont {Z.}~\bibnamefont
				{Wang}}, \bibinfo {author} {\bibfnamefont {M.~P.~A.}\ \bibnamefont {Fisher}},
			\bibinfo {author} {\bibfnamefont {S.~M.}\ \bibnamefont {Girvin}}, \ and\
			\bibinfo {author} {\bibfnamefont {J.~T.}\ \bibnamefont {Chalker}},\ }\href
		{\doibase 10.1103/PhysRevB.61.8326} {\bibfield  {journal} {\bibinfo
				{journal} {Phys. Rev. B}\ }\textbf {\bibinfo {volume} {61}},\ \bibinfo
			{pages} {8326} (\bibinfo {year} {2000})}\BibitemShut {NoStop}%
		\bibitem [{\citenamefont {Pruisken}\ and\ \citenamefont
			{Burmistrov}(2008)}]{PruB08}%
		\BibitemOpen
		\bibfield  {author} {\bibinfo {author} {\bibfnamefont {A.~M.~M.}\
				\bibnamefont {Pruisken}}\ and\ \bibinfo {author} {\bibfnamefont {I.~S.}\
				\bibnamefont {Burmistrov}},\ }\href {\doibase 10.1134/S0021364008040097} {
			\bibfield {journal} {\bibinfo {jhournal} {J. Exp. Theor. Phys. Lett.}\ }
			\textbf {\bibinfo {volume} {87}},\ \bibinfo {pages} {220} (\bibinfo {year}
			{2008})}\BibitemShut {NoStop}
		\bibitem [{\citenamefont {Burmistrov}\ \emph {et~al.}(2011)\citenamefont
			{Burmistrov}, \citenamefont {Bera}, \citenamefont {Evers}, \citenamefont
			{Gornyi},\ and\ \citenamefont {Mirlin}}]{BurBE11}%
		\BibitemOpen
		\bibfield  {author} {\bibinfo {author} {\bibfnamefont {I.~S.}\ \bibnamefont
				{Burmistrov}}, \bibinfo {author} {\bibfnamefont {S.}~\bibnamefont {Bera}},
			\bibinfo {author} {\bibfnamefont {F.}~\bibnamefont {Evers}}, \bibinfo
			{author} {\bibfnamefont {I.~V.}\ \bibnamefont {Gornyi}}, \ and\ \bibinfo
			{author} {\bibfnamefont {A.~D.}\ \bibnamefont {Mirlin}},\ }\href {\doibase
			https://doi.org/10.1016/j.aop.2011.01.005} {\bibfield  {journal} {\bibinfo
				{journal} {Annals of Physics}\ }\textbf {\bibinfo {volume} {326}},\ \bibinfo
			{pages} {1457 } (\bibinfo {year} {2011})}\BibitemShut {NoStop}%
		\bibitem [{Note1()}]{Note1}%
		\BibitemOpen
		\bibinfo {note} {Whereas short-range interactions are irrelevant at the
			noninteracting fixed point, long-range Coulomb interactions are believed to
			be relevant \cite {LeeW96}.}\BibitemShut {Stop}%
		\bibitem [{\citenamefont {Gruzberg}\ \emph {et~al.}(2017)\citenamefont
			{Gruzberg}, \citenamefont {Kl\"umper}, \citenamefont {Nuding},\ and\
			\citenamefont {Sedrakyan}}]{GruKN17}%
		\BibitemOpen
		\bibfield  {author} {\bibinfo {author} {\bibfnamefont {I.~A.}\ \bibnamefont
				{Gruzberg}}, \bibinfo {author} {\bibfnamefont {A.}~\bibnamefont {Kl\"umper}},
			\bibinfo {author} {\bibfnamefont {W.}~\bibnamefont {Nuding}}, \ and\ \bibinfo
			{author} {\bibfnamefont {A.}~\bibnamefont {Sedrakyan}},\ }\href {\doibase
			10.1103/PhysRevB.95.125414} {\bibfield  {journal} {\bibinfo  {journal} {Phys.
					Rev. B}\ }\textbf {\bibinfo {volume} {95}},\ \bibinfo {pages} {125414}
			(\bibinfo {year} {2017})}\BibitemShut {NoStop}%
		\bibitem [{\citenamefont {Ippoliti}\ \emph {et~al.}(2018)\citenamefont
			{Ippoliti}, \citenamefont {Geraedts},\ and\ \citenamefont {Bhatt}}]{IppSB18}%
		\BibitemOpen
		\bibfield  {author} {\bibinfo {author} {\bibfnamefont {M.}~\bibnamefont
				{Ippoliti}}, \bibinfo {author} {\bibfnamefont {S.~D.}\ \bibnamefont
				{Geraedts}}, \ and\ \bibinfo {author} {\bibfnamefont {R.~N.}\ \bibnamefont
				{Bhatt}},\ }\href {\doibase 10.1103/PhysRevB.97.014205} {\bibfield  {journal}
			{\bibinfo  {journal} {Phys. Rev. B}\ }\textbf {\bibinfo {volume} {97}},\
			\bibinfo {pages} {014205} (\bibinfo {year} {2018})}\BibitemShut {NoStop}%
		\bibitem [{\citenamefont {Zhu}\ \emph {et~al.}()\citenamefont {Zhu},
			\citenamefont {Wu}, \citenamefont {Bhatt},\ and\ \citenamefont
			{Wan}}]{ZhuWBW18}%
		\BibitemOpen
		\bibfield  {author} {\bibinfo {author} {\bibfnamefont {Q.}~\bibnamefont
				{Zhu}}, \bibinfo {author} {\bibfnamefont {P.}~\bibnamefont {Wu}}, \bibinfo
			{author} {\bibfnamefont {R.~N.}\ \bibnamefont {Bhatt}}, \ and\ \bibinfo
			{author} {\bibfnamefont {X.}~\bibnamefont {Wan}},\ }\href@noop {} {\ }\Eprint
		{http://arxiv.org/abs/arXiv:1804.00398} {arXiv:1804.00398} \BibitemShut
		{NoStop}%
		\bibitem [{\citenamefont {Peierls}(1933)}]{Pei33}%
		\BibitemOpen
		\bibfield  {author} {\bibinfo {author} {\bibfnamefont {R.~E.}\ \bibnamefont
				{Peierls}},\ }\href@noop {} {\bibfield  {journal} {\bibinfo  {journal} {Z.
					Phys.}\ }\textbf {\bibinfo {volume} {80}},\ \bibinfo {pages} {763–791}
			(\bibinfo {year} {1933})}\BibitemShut {NoStop}%
		\bibitem [{\citenamefont {Luttinger}(1951)}]{Lutt51}%
		\BibitemOpen
		\bibfield  {author} {\bibinfo {author} {\bibfnamefont {J.~M.}\ \bibnamefont
				{Luttinger}},\ }\href@noop {} {\bibfield  {journal} {\bibinfo  {journal}
				{Phys. Rev.}\ }\textbf {\bibinfo {volume} {84}},\ \bibinfo {pages} {814}
			(\bibinfo {year} {1951})}\BibitemShut {NoStop}%
		\bibitem [{\citenamefont {Hofstadter}(1976)}]{Hof76}%
		\BibitemOpen
		\bibfield  {author} {\bibinfo {author} {\bibfnamefont {D.~R.}\ \bibnamefont
				{Hofstadter}},\ }\href {\doibase 10.1103/PhysRevB.14.2239} {\bibfield
			{journal} {\bibinfo  {journal} {Phys. Rev. B}\ }\textbf {\bibinfo {volume}
				{14}},\ \bibinfo {pages} {2239} (\bibinfo {year} {1976})}\BibitemShut
		{NoStop}%
		\bibitem [{\citenamefont {Rammal}(1985)}]{Ram85}%
		\BibitemOpen
		\bibfield  {author} {\bibinfo {author} {\bibfnamefont {R.}~\bibnamefont
				{Rammal}},\ }\href {\doibase 10.1051/jphys:019850046080134500} {\bibfield
			{journal} {\bibinfo  {journal} {J. Phys. France}\ }\textbf {\bibinfo {volume}
				{46}},\ \bibinfo {pages} {1345} (\bibinfo {year} {1985})}\BibitemShut
		{NoStop}%
		\bibitem [{Note2()}]{Note2}%
		\BibitemOpen
		\bibinfo {note} {For irrational $\Phi $, in contrast, the spectrum consists
			of dense but isolated eigenvalues.}\BibitemShut {Stop}%
		\bibitem [{\citenamefont {Schweitzer}\ \emph {et~al.}(1985)\citenamefont
			{Schweitzer}, \citenamefont {Kramer},\ and\ \citenamefont
			{MacKinnon}}]{SchKM85}%
		\BibitemOpen
		\bibfield  {author} {\bibinfo {author} {\bibfnamefont {L.}~\bibnamefont
				{Schweitzer}}, \bibinfo {author} {\bibfnamefont {B.}~\bibnamefont {Kramer}},
			\ and\ \bibinfo {author} {\bibfnamefont {A.}~\bibnamefont {MacKinnon}},\
		}\href {\doibase 10.1007/BF01328845} {\bibfield  {journal} {\bibinfo
				{journal} {Z. Phys. B.}\ }\textbf {\bibinfo {volume} {59}},\ \bibinfo {pages}
			{379} (\bibinfo {year} {1985})}\BibitemShut {NoStop}%
		\bibitem [{\citenamefont {Schweitzer}\ \emph {et~al.}(1984)\citenamefont
			{Schweitzer}, \citenamefont {Kramer},\ and\ \citenamefont
			{MacKinnon}}]{SchKM84}%
		\BibitemOpen
		\bibfield  {author} {\bibinfo {author} {\bibfnamefont {L.}~\bibnamefont
				{Schweitzer}}, \bibinfo {author} {\bibfnamefont {B.}~\bibnamefont {Kramer}},
			\ and\ \bibinfo {author} {\bibfnamefont {A.}~\bibnamefont {MacKinnon}},\
		}\href {http://stacks.iop.org/0022-3719/17/i=23/a=012} {\bibfield  {journal}
			{\bibinfo  {journal} {J. Phys. C Solid State Phys.}\ }\textbf {\bibinfo
				{volume} {17}},\ \bibinfo {pages} {4111} (\bibinfo {year}
			{1984})}\BibitemShut {NoStop}%
		\bibitem [{\citenamefont {Kramer}\ \emph {et~al.}(1984)\citenamefont {Kramer},
			\citenamefont {Schweitzer},\ and\ \citenamefont {MacKinnon}}]{KraSM84}%
		\BibitemOpen
		\bibfield  {author} {\bibinfo {author} {\bibfnamefont {B.}~\bibnamefont
				{Kramer}}, \bibinfo {author} {\bibfnamefont {L.}~\bibnamefont {Schweitzer}},
			\ and\ \bibinfo {author} {\bibfnamefont {A.}~\bibnamefont {MacKinnon}},\
		}\href {\doibase 10.1007/BF01306637} {\bibfield  {journal} {\bibinfo
				{journal} {Z. Phys. B}\ }\textbf {\bibinfo {volume} {56}},\ \bibinfo {pages}
			{297} (\bibinfo {year} {1984})}\BibitemShut {NoStop}%
		\bibitem [{Note3()}]{Note3}%
		\BibitemOpen
		\bibinfo {note} {For numerical stability, we also calculate $\gamma $
			iteratively (in $100$-layer steps) by the algorithm of Refs. \cite
			{SchKM84,KraSM84}.}\BibitemShut {Stop}%
		\bibitem [{Note4()}]{Note4}%
		\BibitemOpen
		\bibinfo {note} {Since the spectrum is symmetric with respect to $E=0$, we
			consider only the positive side of the spectrum, where the lowest LL appears
			on the high energy band edge.}\BibitemShut {Stop}%
		\bibitem [{Note5()}]{Note5}%
		\BibitemOpen
		\bibinfo {note} {See Supplemental Material below for details of the
			simulation parameters, the finite-size corrections at the critical point,
			details of the scaling analysis, and transitions in cases of non-negligible
			intrinsic LL width.}\BibitemShut {Stop}%
		\bibitem [{\citenamefont {Wang}\ \emph {et~al.}(1998)\citenamefont {Wang},
			\citenamefont {Li},\ and\ \citenamefont {Soukoulis}}]{WanLS98}%
		\BibitemOpen
		\bibfield  {author} {\bibinfo {author} {\bibfnamefont {X.}~\bibnamefont
				{Wang}}, \bibinfo {author} {\bibfnamefont {Q.}~\bibnamefont {Li}}, \ and\
			\bibinfo {author} {\bibfnamefont {C.~M.}\ \bibnamefont {Soukoulis}},\
		}\href@noop {} {\bibfield  {journal} {\bibinfo  {journal} {Phys. Rev. B}\
			}\textbf {\bibinfo {volume} {58}},\ \bibinfo {pages} {3576} (\bibinfo {year}
			{1998})}\BibitemShut {NoStop}%
		\bibitem [{\citenamefont {Huckestein}(1994)}]{Huc94}%
		\BibitemOpen
		\bibfield  {author} {\bibinfo {author} {\bibfnamefont {B.}~\bibnamefont
				{Huckestein}},\ }\href {\doibase 10.1103/PhysRevLett.72.1080} {\bibfield
			{journal} {\bibinfo  {journal} {Phys. Rev. Lett.}\ }\textbf {\bibinfo
				{volume} {72}},\ \bibinfo {pages} {1080} (\bibinfo {year}
			{1994})}\BibitemShut {NoStop}%
		\bibitem [{\citenamefont {Janssen}(1998)}]{Jan98}%
		\BibitemOpen
		\bibfield  {author} {\bibinfo {author} {\bibfnamefont {M.}~\bibnamefont
				{Janssen}},\ }\href@noop {} {\bibfield  {journal} {\bibinfo  {journal} {Phys.
					Rep.}\ }\textbf {\bibinfo {volume} {295}},\ \bibinfo {pages} {1} (\bibinfo
			{year} {1998})}\BibitemShut {NoStop}%
		\bibitem [{\citenamefont {Evers}\ \emph {et~al.}(2008)\citenamefont {Evers},
			\citenamefont {Mildenberger},\ and\ \citenamefont {Mirlin}}]{EveMM08}%
		\BibitemOpen
		\bibfield  {author} {\bibinfo {author} {\bibfnamefont {F.}~\bibnamefont
				{Evers}}, \bibinfo {author} {\bibfnamefont {A.}~\bibnamefont {Mildenberger}},
			\ and\ \bibinfo {author} {\bibfnamefont {A.~D.}\ \bibnamefont {Mirlin}},\
		}\href {\doibase 10.1103/PhysRevLett.101.116803} {\bibfield  {journal}
			{\bibinfo  {journal} {Phys. Rev. Lett.}\ }\textbf {\bibinfo {volume} {101}},\
			\bibinfo {pages} {116803} (\bibinfo {year} {2008})}\BibitemShut {NoStop}%
		\bibitem [{\citenamefont {Obuse}\ \emph {et~al.}(2008)\citenamefont {Obuse},
			\citenamefont {Subramaniam}, \citenamefont {Furusaki}, \citenamefont
			{Gruzberg},\ and\ \citenamefont {Ludwig}}]{ObuSF08}%
		\BibitemOpen
		\bibfield  {author} {\bibinfo {author} {\bibfnamefont {H.}~\bibnamefont
				{Obuse}}, \bibinfo {author} {\bibfnamefont {A.~R.}\ \bibnamefont
				{Subramaniam}}, \bibinfo {author} {\bibfnamefont {A.}~\bibnamefont
				{Furusaki}}, \bibinfo {author} {\bibfnamefont {I.~A.}\ \bibnamefont
				{Gruzberg}}, \ and\ \bibinfo {author} {\bibfnamefont {A.~W.~W.}\ \bibnamefont
				{Ludwig}},\ }\href {\doibase 10.1103/PhysRevLett.101.116802} {\bibfield
			{journal} {\bibinfo  {journal} {Phys. Rev. Lett.}\ }\textbf {\bibinfo
				{volume} {101}},\ \bibinfo {pages} {116802} (\bibinfo {year}
			{2008})}\BibitemShut {NoStop}%
		\bibitem [{Note6()}]{Note6}%
		\BibitemOpen
		\bibinfo {note} {The lattice in the CC network plays a very different role
			than in our work. In the CC network, it represents the semiclassical electron
			path whereas in our work, it sets up the Hamiltonian while the electron
			motion is irregular.}\BibitemShut {Stop}%
		\bibitem [{Note7()}]{Note7}%
		\BibitemOpen
		\bibinfo {note} {It is remarkable that the irrelevant exponent $y=4.3(2)$
			observed in Ref.\ \cite {ZhuWBW18}, expected to be universal, is much larger
			than those found in the current work or in other recent investigations based
			on CC network models.}\BibitemShut {Stop}%
		\bibitem [{Note8()}]{Note8}%
		\BibitemOpen
		\bibinfo {note} {It is worth emphasizing that simple manual scaling plots are
			not able to resolve the slowly decaying corrections in this
			problem.}\BibitemShut {Stop}%
	\end{thebibliography}

\begin{thebibliography}{6}%
	\makeatletter
	\providecommand \@ifxundefined [1]{%
		\@ifx{#1\undefined}
	}%
	\providecommand \@ifnum [1]{%
		\ifnum #1\expandafter \@firstoftwo
		\else \expandafter \@secondoftwo
		\fi
	}%
	\providecommand \@ifx [1]{%
		\ifx #1\expandafter \@firstoftwo
		\else \expandafter \@secondoftwo
		\fi
	}%
	\providecommand \natexlab [1]{#1}%
	\providecommand \enquote  [1]{``#1''}%
	\providecommand \bibnamefont  [1]{#1}%
	\providecommand \bibfnamefont [1]{#1}%
	\providecommand \citenamefont [1]{#1}%
	\providecommand \href@noop [0]{\@secondoftwo}%
	\providecommand \href [0]{\begingroup \@sanitize@url \@href}%
	\providecommand \@href[1]{\@@startlink{#1}\@@href}%
	\providecommand \@@href[1]{\endgroup#1\@@endlink}%
	\providecommand \@sanitize@url [0]{\catcode `\\12\catcode `\$12\catcode
		`\&12\catcode `\#12\catcode `\^12\catcode `\_12\catcode `\%12\relax}%
	\providecommand \@@startlink[1]{}%
	\providecommand \@@endlink[0]{}%
	\providecommand \url  [0]{\begingroup\@sanitize@url \@url }%
	\providecommand \@url [1]{\endgroup\@href {#1}{\urlprefix }}%
	\providecommand \urlprefix  [0]{URL }%
	\providecommand \Eprint [0]{\href }%
	\providecommand \doibase [0]{http://dx.doi.org/}%
	\providecommand \selectlanguage [0]{\@gobble}%
	\providecommand \bibinfo  [0]{\@secondoftwo}%
	\providecommand \bibfield  [0]{\@secondoftwo}%
	\providecommand \translation [1]{[#1]}%
	\providecommand \BibitemOpen [0]{}%
	\providecommand \bibitemStop [0]{}%
	\providecommand \bibitemNoStop [0]{.\EOS\space}%
	\providecommand \EOS [0]{\spacefactor3000\relax}%
	\providecommand \BibitemShut  [1]{\csname bibitem#1\endcsname}%
	\let\auto@bib@innerbib\@empty
	%</preamble>
	\bibitem [{\citenamefont {Slevin}\ and\ \citenamefont
		{Ohtsuki}(2009)}]{SleO09S}%
	\BibitemOpen
	\bibfield  {author} {\bibinfo {author} {\bibfnamefont {K.}~\bibnamefont
			{Slevin}}\ and\ \bibinfo {author} {\bibfnamefont {T.}~\bibnamefont
			{Ohtsuki}},\ }\href {\doibase 10.1103/PhysRevB.80.041304} {\bibfield
		{journal} {\bibinfo  {journal} {Phys. Rev. B}\ }\textbf {\bibinfo {volume}
			{80}},\ \bibinfo {pages} {041304} (\bibinfo {year} {2009})}\BibitemShut
	{NoStop}%
	\bibitem [{\citenamefont {Nuding}\ \emph {et~al.}(2015)\citenamefont {Nuding},
		\citenamefont {Kl\"umper},\ and\ \citenamefont {Sedrakyan}}]{NudKS15S}%
	\BibitemOpen
	\bibfield  {author} {\bibinfo {author} {\bibfnamefont {W.}~\bibnamefont
			{Nuding}}, \bibinfo {author} {\bibfnamefont {A.}~\bibnamefont {Kl\"umper}}, \
		and\ \bibinfo {author} {\bibfnamefont {A.}~\bibnamefont {Sedrakyan}},\ }\href
	{\doibase 10.1103/PhysRevB.91.115107} {\bibfield  {journal} {\bibinfo
			{journal} {Phys. Rev. B}\ }\textbf {\bibinfo {volume} {91}},\ \bibinfo
		{pages} {115107} (\bibinfo {year} {2015})}\BibitemShut {NoStop}%
	\bibitem [{\citenamefont {Obuse}\ \emph {et~al.}(2012)\citenamefont {Obuse},
		\citenamefont {Gruzberg},\ and\ \citenamefont {Evers}}]{ObuGE12S}%
	\BibitemOpen
	\bibfield  {author} {\bibinfo {author} {\bibfnamefont {H.}~\bibnamefont
			{Obuse}}, \bibinfo {author} {\bibfnamefont {I.~A.}\ \bibnamefont {Gruzberg}},
		\ and\ \bibinfo {author} {\bibfnamefont {F.}~\bibnamefont {Evers}},\ }\href
	{\doibase 10.1103/PhysRevLett.109.206804} {\bibfield  {journal} {\bibinfo
			{journal} {Phys. Rev. Lett.}\ }\textbf {\bibinfo {volume} {109}},\ \bibinfo
		{pages} {206804} (\bibinfo {year} {2012})}\BibitemShut {NoStop}%
	\bibitem [{\citenamefont {Gruzberg}\ \emph {et~al.}(2017)\citenamefont
		{Gruzberg}, \citenamefont {Kl\"umper}, \citenamefont {Nuding},\ and\
		\citenamefont {Sedrakyan}}]{GruKN17S}%
	\BibitemOpen
	\bibfield  {author} {\bibinfo {author} {\bibfnamefont {I.~A.}\ \bibnamefont
			{Gruzberg}}, \bibinfo {author} {\bibfnamefont {A.}~\bibnamefont {Kl\"umper}},
		\bibinfo {author} {\bibfnamefont {W.}~\bibnamefont {Nuding}}, \ and\ \bibinfo
		{author} {\bibfnamefont {A.}~\bibnamefont {Sedrakyan}},\ }\href {\doibase
		10.1103/PhysRevB.95.125414} {\bibfield  {journal} {\bibinfo  {journal} {Phys.
				Rev. B}\ }\textbf {\bibinfo {volume} {95}},\ \bibinfo {pages} {125414}
		(\bibinfo {year} {2017})}\BibitemShut {NoStop}%
	\bibitem [{\citenamefont {Slevin}\ and\ \citenamefont
		{Ohtsuki}(2012)}]{SLEO12S}%
	\BibitemOpen
	\bibfield  {author} {\bibinfo {author} {\bibfnamefont {K.}~\bibnamefont
			{Slevin}}\ and\ \bibinfo {author} {\bibfnamefont {T.}~\bibnamefont
			{Ohtsuki}},\ }\href {\doibase 10.1142/S2010194512006162} {
		\bibfield {journal} {\bibinfo {journal} {Int. J. Mod. Phys. Conf. Ser.}\ }
		\textbf	{\bibinfo {volume} {11}},\ 
		\bibinfo {pages} {60} (\bibinfo {year} {2012})}
	\BibitemShut {NoStop}%
	\bibitem [{\citenamefont {Amado}\ \emph {et~al.}(2011)\citenamefont {Amado},
		\citenamefont {Malyshev}, \citenamefont {Sedrakyan},\ and\ \citenamefont
		{Dominguez-Adame}}]{AmaMS11S}%
	\BibitemOpen
	\bibfield  {author} {\bibinfo {author} {\bibfnamefont {M.}~\bibnamefont
			{Amado}}, \bibinfo {author} {\bibfnamefont {A.~V.}\ \bibnamefont {Malyshev}},
		\bibinfo {author} {\bibfnamefont {A.}~\bibnamefont {Sedrakyan}}, \ and\
		\bibinfo {author} {\bibfnamefont {F.}~\bibnamefont {Dominguez-Adame}},\
	}\href {\doibase 10.1103/PhysRevLett.107.066402} {\bibfield  {journal}
		{\bibinfo  {journal} {Phys. Rev. Lett.}\ }\textbf {\bibinfo {volume} {107}},\
		\bibinfo {pages} {066402} (\bibinfo {year} {2011})}\BibitemShut {NoStop}%
\end{thebibliography}
\end{document}